\documentclass[final,3p,times,fleqn]{elsarticle}
\usepackage{mathtools,booktabs}
\usepackage{bm}
\usepackage{autobreak}
\usepackage{amsmath}
\usepackage{amsthm}
\usepackage{gensymb}
\usepackage{subfigure}
\usepackage{epstopdf}
\usepackage{color}
\usepackage{hyperref}
\newtheorem{remark}{Remark}
\biboptions{sort&compress}
\begin{document}
\begin{frontmatter}
	\title{Consistent and conservative phase-field based lattice Boltzmann method for incompressible two-phase flows}
	\author[a]{Chengjie Zhan}
	%\ead{zhancj_@hust.edu.cn}
	\author[a,b]{Zhenhua Chai \corref{cor1}}
	\ead{hustczh@hust.edu.cn}
	\author[a,b]{Baochang Shi}
	%\ead{hustczh@hust.edu.cn}
	\address[a]{School of Mathematics and Statistics, Huazhong University of Science and Technology, Wuhan, 430074, China}
	\address[b]{Hubei Key Laboratory of Engineering Modeling and Scientific Computing, Huazhong University of Science and Technology, Wuhan, 430074, China}
	\cortext[cor1]{Corresponding author.}
\begin{abstract}
	In this work, we consider a general consistent and conservative phase-field model for the incompressible two-phase flows. In this model, not only the Cahn-Hilliard or Allen-Cahn equation can be adopted, but also the mass and the momentum fluxes in the Navier-Stokes equations are reformulated such that the consistency of reduction, consistency of mass and momentum transport, and the consistency of mass conservation are satisfied. We further develop a lattice Boltzmann (LB) method and show that through the direct Taylor expansion, the present LB method can correctly recover the consistent and conservative phase-field model. Additionally, if the divergence of the extra momentum flux is seen as a force term, the extra force in the present LB method would include another term which has not been considered in the previous LB models. To quantitatively evaluate the incompressibility and the consistency of the mass conservation, two statistical variables are introduced in the study of the deformation of a square droplet, and the results show that the present LB method is more accurate. The layered Poiseuille flow and a droplet spreading on an ideal wall are further investigated, and the numerical results are in good agreement with the analytical solutions. Finally, the problems of the Rayleigh-Taylor instability and dam break with the high Reynolds numbers and/or large density ratios are studied, and it is found that the present consistent and conservative LB method is robust for such complex two-phase flows. 
\end{abstract}
\end{frontmatter}
\section{Introduction}
Two-phase flows are ubiquitous both in nature \cite{Gan2009JMM} and engineering \cite{Li2005JCIS,Teh2008LC}, and have attracted much attention due to a wide range of applications and the complicated physical characteristics.
The phase-field model, as a diffuse interface approach, has usually been used in the study of two-phase flows \cite{Anderson1998ARFM,Badalassi2003JCP,Ding2007JCP,Chiu2011JCP}. Although the physical thickness of the interface in the phase-field model is impossible to numerically resolve the interface of realistic immiscible two-phase flows, it still has some distinct advantages \cite{Mirjalili2021JCP}, for instance, the implementation of the phase-field model does not need to distinguish different cells, the normal vectors and curvature values can be computed directly from the phase field depicted by the order parameter, and the mass of system is conservative. 
The popularly used governing equation of the phase field is the Cahn-Hilliard (CH) equation \cite{Cahn1996EJAM}, in which there is a fourth-order spatial derivative term. Recently, the conservative Allen-Cahn (AC) equation \cite{Sun2007JCP,Chiu2011JCP} has been another strategy to capture the phase interface, which is simpler than the CH equation since only a second-order spatial derivative is included.  

However, it should be noted that when we consider the phase-field method for the incompressible two-phase flows, the mass conservation equation $\partial_t\rho+\nabla\cdot\left(\rho\mathbf{u}\right)=0$ is no longer a consequence of the incompressibility condition \cite{Shen2011LNS}, and is inconsistent with the phase-field model, which may also cause some numerical instability in the study of the problems with high Reynolds numbers and/or large density ratios. To keep the governing equations consistent with each other, Huang et al. \cite{Huang2020JCPa,Huang2020JCPb} proposed three consistency conditions, i.e., consistency of reduction, consistency of mass and momentum transport, and consistency of mass conservation. To satisfy these consistency conditions, the mass flux in the Navier-Stokes (NS) equations should be reformulated according to the phase-field model. In this case, one can get the consistent and conservative phase-field-NS equations for incompressible two-phase flows. Almost as the same time, the same consistent model was also developed by Mirjalili and Mani from a similar point of view \cite{Mirjalili2021JCP}. 

Lattice Boltzmann (LB) method, as a mesoscopic numerical approach, has been developed into an efficient tool in the study of the complex fluid systems \cite{Higuera1989EPL,Benzi1992PR,Qian1995ARCP,Chen1998ARFM,Aidun2010ARFM} and non-linear partial differential equations \cite{Dawson1993JCP,Blaak2000CPC,Shi2009PRE,Chai2013PRE,Zhao2020CMA} during the past three decades. Due to its kinetic background \cite{Succi2001,Kruger2017}, the LB method has some distinct features, including the easy implementation of boundary conditions and fully parallel algorithm \cite{Chen1998ARFM}. Considering the advantages of phase-field model and LB method, some phase-field based LB models have been developed.
He et al. \cite{He1999JCP} first proposed a phase-field LB model for incompressible two-phase flows. However, as point out in Ref. \cite{Zheng2005PRE}, the recovered macroscopic equation is different from the CH equation in phase-field theory. 
To derive the correct CH equation, Zu and He \cite{Zu2013PRE} developed another LB model through introducing a spatial difference term of the equilibrium distribution function in the evolution equation.  
Liang et al. \cite{Liang2014PRE} also designed another LB model for the CH equation, in which a time derivative term included in the evolution function to remove the additional terms in the recovered equation.
On the other hand, two LB models were independently proposed for AC equation, and a comparative study of the LB models for AC and CH equations was performed in Refs. \cite{Ren2016PRE,Wang2016PRE}.
Liang et al. \cite{Liang2018PRE} further presented a simple LB model with AC phase-field theory for two-phase flows, and found that the model is more accurate than the previous LB model \cite{Ren2016PRE}, especially for the problems with large density ratios.
Recently, Yuan et al. \cite{Yuan2020CMA} proposed a generalized LB model for both incompressible and nearly incompressible NS equations, and in the framework of this model, a new phase-field-based LB model is developed for two-phase flows. 
However, all above LB models only focus on the system with mass flux $\rho\mathbf{u}$, which does not meet the consistency of mass conservation \cite{Huang2020JCPa,Huang2020JCPb} since the flux caused by the diffusion in the interface layer is not included. To overcome this problem, in this work a consistent and conservative phase-field based LB method is proposed for the incompressible two-phase flows, which are depicted by the consistent and conservative phase-field-NS equations.  
In this method, we first present a general LB model for the phase-field equation that can be viewed as a general form of the CH and AC equations considered in the previous works \cite{Liang2014PRE,Wang2016PRE,Wang2019Capillarity}, and then develop a new LB model for the consistent and conservative NS equations.

The rest of this paper is organized as follows. The governing equations and the consistency conditions are first given in Section \ref{GoverningEq}. Then in Section \ref{LBmodel}, a general multiple-relaxation-time (MRT) LB method for the consistent and conservative phase-field-NS equations is proposed, which can recover the macroscopic equations through the direct Taylor expansion analysis. 
In Section \ref{numerical}, several typical benchmark problems, including the deformation of a square droplet, the layered Poiseuille flow, a droplet spreading on an ideal wall, the Rayleigh-Taylor instability (RTI) and dam break, are used to test the developed LB method. The results show that the present LB method is more accurate and more stable, especially for the two-phase flows with the high Reynolds numbers and/or large density ratios. Finally, some conclusions are summarized in Section \ref{conclusion}.

\section{Governing equations and the consistency conditions}\label{GoverningEq}
The general phase-field model can be expressed by
\begin{equation}\label{PhaseField}
	\frac{\partial\phi}{\partial t}+\nabla\cdot\left(\phi\mathbf{u}\right)=\nabla\cdot\left[M_{\phi}\mathbf{D}\left(\phi\right)\right]+S_{\phi},
\end{equation}
where $\phi$ is the order parameter, and is dependent on the density of the fluid. $\mathbf{u}$ is the fluid velocity, $M_{\phi}$ is the mobility, $\mathbf{D}\left(\phi\right)$ is related to the diffusion flux, and $S_{\phi}$ is the source term which should be zero for the immiscible fluids. Actually, the classical CH and AC equations can be considered as the specific forms of Eq.\,(\ref{PhaseField}) once the particular $\mathbf{D}\left(\phi\right)$ are chosen (see some details below). 

In phase-field theory, the free energy density of a system can be written as \cite{Shen2011LNS,Lee2012CPC,Jacqmin1999JCP}
\begin{equation}
	f\left(\phi,\nabla\phi\right)=\psi\left(\phi\right)+\frac{k}{2}|\nabla\phi|^2,
\end{equation}
where $\psi\left(\phi\right)=\beta\left(\phi-\phi_A\right)^2\left(\phi-\phi_B\right)^2$ is the bulk energy \cite{Shen2011LNS,Lee2012CPC}, $\phi_A$ and $\phi_B$ are also two constants, and for simplicity, it is assumed that $\phi_A>\phi_B$. $\beta$ and $k$ are two physical parameters related to the interfacial thickness $W$ and the surface tension $\sigma$,
\begin{equation}
	\beta=\frac{12\sigma}{W\left(\phi_A-\phi_B\right)^4},\quad k=\frac{3}{2}\frac{\sigma W}{\left(\phi_A-\phi_B\right)^2}.
\end{equation}
From the free energy, we can determine the mixing energy $F$ and chemical potential $\mu_{\phi}$,
\begin{equation}
	F\left(\phi,\nabla\phi\right)=\int_{\Omega}f\left(\phi,\nabla\phi\right)d\Omega=\int_{\Omega}\left[\psi\left(\phi\right)+\frac{k}{2}|\nabla\phi|^2\right]d\Omega,
\end{equation}
\begin{equation}
	\mu_{\phi}=\frac{\delta F}{\delta \phi}=\psi'\left(\phi\right)-k\nabla^2\phi=4\beta\left(\phi-\phi_A\right)\left(\phi-\phi_B\right)\left(\phi-\frac{\phi_A+\phi_B}{2}\right)-k\nabla^2\phi.	
\end{equation}

With above order parameter $\phi$ and chemical potential $\mu_{\phi}$, we can derive the classical CH and AC equations from Eq.\,(\ref{PhaseField}) by choosing different form of $\mathbf{D}\left(\phi\right)$.

\noindent\uppercase\expandafter{\romannumeral1}: $\mathbf{D}\left(\phi\right)=\nabla\mu_{\phi}$ gives the fourth-order CH equation,
\begin{equation}\label{CHE}
	\frac{\partial\phi}{\partial t}+\nabla\cdot\left(\phi\mathbf{u}\right)=\nabla\cdot M_{\phi}\nabla\mu_{\phi}.
\end{equation}

\noindent\uppercase\expandafter{\romannumeral2}: $\mathbf{D}\left(\phi\right)=\nabla\phi-\lambda\mathbf{n}$ leads to the second-order AC equation,
\begin{equation}\label{ACE}
	\frac{\partial\phi}{\partial t}+\nabla\cdot\left(\phi\mathbf{u}\right)=\nabla\cdot M_{\phi}\left(\nabla\phi-\lambda\mathbf{n}\right),
\end{equation}
where $\lambda$ and $\mathbf{n}$ are defined by
\begin{equation}
	\lambda=\sqrt{\frac{2\beta}{k}}\left(\phi_A-\phi\right)\left(\phi-\phi_B\right),\quad \mathbf{n}=\frac{\nabla\phi}{|\nabla\phi|}.
\end{equation}
Actually, under the no-flux boundary condition, one can obtain $d\phi/d t=0$, which means the system is conservative.

To describe the fluid flows, the following incompressible NS equations are used \cite{Huang2020JCPa,Huang2020JCPb},
\begin{subequations}\label{NS}
	\begin{equation}\label{NSrho}
		\frac{\partial\rho}{\partial t}+\nabla\cdot\mathbf{m}=S_m,
	\end{equation}
	\begin{equation}\label{NS1}
		\nabla\cdot\mathbf{u}=0,
	\end{equation}
	\begin{equation}\label{NS2}
		\frac{\partial\left(\rho\mathbf{u}\right)}{\partial t}+\nabla\cdot\left(\mathbf{mu}\right)=-\nabla p+\nabla\cdot\mu\left[\nabla\mathbf{u}+\left(\nabla\mathbf{u}\right)^T\right]+\mathbf{F}_s+\mathbf{S}_{\mathbf{u}},
	\end{equation}
\end{subequations}
where $\rho$ is the fluid density, $\mathbf{m}$ is the mass flux, $S_m$ is the mass source, $p$ is the hydrodynamic pressure, $\mu=\rho\nu$ is the dynamic viscosity with $\nu$ representing the kinematic viscosity, $\mathbf{F}_s=\mu_{\phi}\nabla\phi$ is the surface force, and $\mathbf{S}_{\mathbf{u}}$ is the momentum source or other external force. In above equations, Eq.\,(\ref{NSrho}) denotes the mass conservation, Eq.\,(\ref{NS1}) means the incompressibility or volume conservation, and Eq.\,(\ref{NS2}) represents the conversation of momentum. 

The distribution of fluid density in a two-phase system is physically consistent with that of the order parameter, and can be given as 
\begin{equation}\label{rho-phi}
	\rho=\frac{d\rho}{d\phi}\left(\phi-\phi_B\right)+\rho_B,
\end{equation}
where $\frac{d\rho}{d\phi}=\left(\rho_A-\rho_B\right)/\left(\phi_A-\phi_B\right)$ with $\rho_A$ and $\rho_B$ being the densities of fluid A and fluid B, respectively.
 
In the phase-field-NS system, to ensure the viscosity change smoothly across the interface, several expressions have been adopted. The first is that the viscosity is assumed to be a linear function of the order parameter,
\begin{equation}\label{linearMu}
	\mu=\frac{\phi-\phi_B}{\phi_A-\phi_B}\left(\mu_A-\mu_B\right)+\mu_B,
\end{equation}
where $\mu_A$ and $\mu_B$ are the dynamic viscosities of the two different phases.  
The second is that the viscosity is expressed as an inverse linear form,
\begin{equation}\label{inverseNu}
	\frac{1}{\nu}=\frac{\phi-\phi_B}{\phi_A-\phi_B}\left(\frac{1}{\nu_A}-\frac{1}{\nu_B}\right)+\frac{1}{\nu_B},
\end{equation}
where $\nu_A$ and $\nu_B$ are the kinematic viscosities of the two phases. We note that the first linear form is very popular for its simplicity \cite{He1999JCP,Huang2020JCPa}, while the second one seems more accurate or stability in the study of two-phase flows with the high Reynolds numbers and large density ratios. In this work, without otherwise stated, the first one would be adopted.

To give a consistent mathematical model, Huang et al. \cite{Huang2020JCPa,Huang2020JCPb} considered the following consistency conditions, 
\begin{itemize}
	\item \textbf{Consistency of reduction}: When $\rho_A=\rho_B$ and $\mu_A=\mu_B$ [or $\phi\equiv\phi_A(\phi_B)$], the momentum conservation equation (\ref{NS2}) in two-phase flow model should reduce to the  single-phase one, i.e., $\mathbf{m}=\rho\mathbf{u}$ and $\nabla\cdot\left[\mu\left(\nabla\mathbf{u}\right)^T\right]=0$.
	\item \textbf{Consistency of mass and momentum transport}: The momentum flux in the conservation equation (\ref{NS2}) should be consistently computed from the mass flux obtained from the conservation equation (\ref{NSrho}).
	\item \textbf{Consistency of mass conservation}: The mass conservation equation (\ref{NSrho}) should be consistent with the phase-field model (\ref{PhaseField}) based on the relation (\ref{rho-phi}). The mass flux $\mathbf{m}$ in the mass conservation equation (\ref{NSrho}) should lead to a zero mass source, i.e., $S_m=0$.
\end{itemize}
Here it should be noted that if the source term $S_m=0$ and the mass flux is defined as the usual form $\mathbf{m}=\rho\mathbf{u}$, the commonly used NS equations cannot preserve the consistency of mass conservation, which may lead to some nonphysical solutions.

Based on the analysis in the previous works \cite{Huang2020JCPa,Huang2020JCPb}, in order to obtain the physical solutions, one needs to define a consistent mass flux $\mathbf{m}$ (hereafter $\mathbf{m}^{CC}$) to replace $\rho\mathbf{u}$ such that the consistency of mass conservation with $S_m=0$ can be guaranteed. Actually, through multiplying $d\rho/d\phi$ on the both sides of Eq.\,(\ref{PhaseField}), we have
\begin{equation}\label{PFPrho}
	\frac{d\rho}{d\phi}\frac{\partial\phi}{\partial t}+\frac{d\rho}{d\phi}\nabla\cdot\left(\phi\mathbf{u}\right)=\frac{d\rho}{d\phi}\nabla\cdot\left[M_{\phi}\mathbf{D}\left(\phi\right)\right].
\end{equation}
According to the relation (\ref{rho-phi}) and the incompressible condition $\nabla\cdot\mathbf{u}=0$, one can get the following conservative form of the mass conservation equation,
\begin{equation}\label{PFrho}
	\frac{\partial\rho}{\partial t}+\nabla\cdot\left[\rho\mathbf{u}-\frac{d\rho}{d\phi}M_{\phi}\mathbf{D}\left(\phi\right)\right]=0,
\end{equation}
from which we can determine the consistent mass flux $\mathbf{m}^{CC}$, 
\begin{equation}
	\mathbf{m}^{CC}=\rho\mathbf{u}+\mathbf{m}^{\phi C}, \quad \mathbf{m}^{\phi C}=-\frac{d\rho}{d\phi}M_{\phi}\mathbf{D}\left(\phi\right),
\end{equation}
where $\mathbf{m}^{\phi C}$ denotes the mass diffusion between different phases. We would like to point out that one do not need to solve the mass conservation equation (\ref{PFrho}) because of its consistency with phase-field model, and meanwhile, with above definition of consistent mass flux, all the three consistency conditions mentioned above can be satisfied.

\section{The consistent and conservative lattice Boltzmann method for  incompressible two-phase flows}\label{LBmodel}
In this section, we will develop a consistent and conservative LB method for the following coupled system,
\begin{subequations}
	\begin{equation}\label{PF}
		\frac{\partial\phi}{\partial t}+\nabla\cdot\left(\phi\mathbf{u}\right)=\nabla\cdot\left[M_{\phi}\mathbf{D}\left(\phi\right)\right],
	\end{equation}
	\begin{equation}\label{NS1-1}
		\nabla\cdot\mathbf{u}=0,
	\end{equation}
	\begin{equation}\label{NS2-mphiC}
		\frac{\partial\left(\rho\mathbf{u}\right)}{\partial t}+\nabla\cdot\left(\rho\mathbf{uu}+\mathbf{m}^{\phi C}\mathbf{u}\right)=-\nabla p+\nabla\cdot\mu\left[\nabla\mathbf{u}+\left(\nabla\mathbf{u}\right)^T\right]+\mathbf{F},
	\end{equation}
\end{subequations}
where $\mathbf{F}$ is the total external force. The term $\nabla\cdot\left(\mathbf{m}^{\phi C}\mathbf{u}\right)$ can be seen as a force term and placed on the right hand side of Eq.\,(\ref{NS2-mphiC}), but it is different from the additional interfacial force in Refs. \cite{Li2012PRE,Ren2016PRE} (see details in \ref{comparison}). 

We first consider the LB model for phase-field equation, which is a general form of the CH and AC equations adopted in the previous works \cite{Liang2014PRE,Wang2016PRE,Wang2019Capillarity}. In this model, the evolution equation is written as \cite{Chai2020PRE}
\begin{equation}\label{evoPF}
	\begin{aligned}
		f_i(\mathbf{x}+\mathbf{c}_i\Delta t,t+\Delta t)=&f_i(\mathbf{x},t)-\left(\mathbf{M}^{-1}\mathbf{S}_f\mathbf{M}\right)_{ij}\left[f_j(\mathbf{x},t)-f_j^{eq}(\mathbf{x},t)\right]+\Delta t\left[\mathbf{M}^{-1}\left(\mathbf{I}-\mathbf{S}/2\right)\mathbf{M}\right]_{ij}R_j(\mathbf{x},t)\\
		=&\mathbf{M}_{ij}^{-1}\left[\left(\mathbf{I}-\mathbf{S}_f\right)\mathbf{m}_f+\mathbf{S}_f\mathbf{m}_f^{eq}+\Delta t\left(\mathbf{I}-\mathbf{S}_f/2\right)\mathbf{m}_R\right]_j,
	\end{aligned}	
\end{equation}
where $f_i\left(\mathbf{x},t\right)$ ($i=0,1,\cdots,q-1$ with $q$ being the number of discrete velocity directions) represents the distribution function of order parameter $\phi$ at position $\mathbf{x}$ and time $t$, and $f_i^{eq}\left(\mathbf{x},t\right)$ is the corresponding equilibrium distribution function \cite{Liang2014PRE,Wang2016PRE,Wang2019Capillarity}. $\mathbf{c}_i$ is the discrete velocity, $\Delta t$ is the time step. $\mathbf{M}$ is a $q\times q$ transformation matrix, $\mathbf{m}_f=\left(\mathbf{M}_{ij}f_j\right)$, $\mathbf{m}_f^{eq}=\left(\mathbf{M}_{ij}f_j^{eq}\right)$,  $\mathbf{m}_R=\left(\mathbf{M}_{ij}R_j\right)$,  and $\mathbf{S}_f=\mathbf{diag}\left(s_f^0, s_f^1, s_f^2, \cdots,s_f^{q-1}\right)$ is the diagonal relaxation matrix. With the direct Taylor expansion, Eq.\,(\ref{PF}) can be recovered correctly.

In the following, we will develop a new LB model for the consistent and conservative incompressible NS equations. The evolution equation of this model reads	
\begin{equation}\label{evoNS}
	g_i(\mathbf{x}+\mathbf{c}_i\Delta t,t+\Delta t)=g_i(\mathbf{x},t)-\Lambda_{ij}\left[g_j(\mathbf{x},t)-g_j^{eq}(\mathbf{x},t)\right]+\Delta t\left(\delta_{ij}-\frac{\Lambda_{ij}}{2}\right)F_j(\mathbf{x},t),
\end{equation}
where $g_i\left(\mathbf{x},t\right)$ is the distribution function of the fluid field, $\bm{\Lambda}=\left(\Lambda_{ij}\right)$ represents the invertible collision matrix. 

To obtain the incompressible NS equations (\ref{NS1-1})-(\ref{NS2-mphiC}), the equilibrium distribution function $g_i^{eq}\left(\mathbf{x},t\right)$ is designed as
\begin{equation}\label{geq}
	g_i^{eq}=\alpha_i+s_i(\phi,\mathbf{u}),
\end{equation}
where $\alpha_0=\left(\omega_0-1\right)p/c_s^2+\rho_0$ with $\rho_0$ being the constant density, $\alpha_i=\omega_ip/c_s^2 (i\neq0)$, and $s_i(\phi,\mathbf{u})$ is given by
\begin{equation}
	s_i(\phi,\mathbf{u})=\omega_i\left[\frac{\mathbf{c}_i\cdot\left(\rho\mathbf{u}\right)}{c_s^2}+\frac{\left(\rho\mathbf{uu}+\mathbf{m}^{\phi C}\mathbf{u}\right):\left(\mathbf{c}_i\mathbf{c}_i-c_s^2\mathbf{I}\right)}{2c_s^4}\right],
\end{equation}
where $\omega_i$ is the weight coefficient, $c_s$ represents the lattice sound speed. The distribution function of the force term $F_i\left(\mathbf{x},t\right)$ is defined as
\begin{equation}
	F_i=\omega_i\left[\mathbf{u}\cdot\nabla\rho+\frac{\mathbf{c}_i\cdot\mathbf{F}}{c_s^2}+\frac{\left(\mathbf{M}_{2F}-c_s^2\mathbf{u}\cdot\nabla\rho\mathbf{I}\right):(\mathbf{c}_i\mathbf{c}_i-c_s^2\mathbf{I})}{2c_s^4}\right],
\end{equation}
the expression of $\mathbf{M}_{2F}$ will be given in Section \ref{Taylorex}.

The order parameter, macroscopic velocity, and pressure are computed by
\begin{subequations}
	\begin{equation}
		\phi=\sum_{i}f_i,
	\end{equation}
	\begin{equation}\label{comU}
		\rho\mathbf{u}=\sum_{i}\mathbf{c}_ig_i+\frac{\Delta t}{2}\mathbf{F},
	\end{equation}
	\begin{equation}\label{comP}
		p=\frac{c_s^2}{1-\omega_0}\left[\sum_{i\neq0}g_i+\left(\frac{1}{2}+H\right)\Delta t\mathbf{u}\cdot\nabla\rho+s_0\left(\phi,\mathbf{u}\right)+Jd\Delta t\partial_tp+K\Delta t\partial_t\left(\rho\mathbf{u}\cdot\mathbf{u}+\mathbf{m}^{\phi C}\cdot\mathbf{u}\right)\right],
	\end{equation}
\end{subequations}
where $d$ is the dimension of space, $H$, $J$ and $K$ are parameters which can be determined by a specified collision martix $\bm{\Lambda}$ (see details in Section \ref{computingP}).

\subsection{The direct Taylor expansion of present LB model for incompressible Navier-Stokes equations}\label{Taylorex}
Applying the Taylor expansion to the left hand side of Eq.\,(\ref{evoNS}), we have \cite{Chai2020PRE}
\begin{equation}\label{taylorEx}
	\sum_{l=1}^{N}\frac{\Delta t^{l}}{l!}D_{i}^{l}g_i+O(\Delta t^{N+1})=-\Lambda_{ij}\left(g_j-g_j^{eq}\right)+\Delta t\left(\delta_{ij}-\frac{\Lambda_{ij}}{2}\right)F_j,
\end{equation}
where $D_i=\partial_t+\mathbf{c}_i\cdot\nabla$. Based on above equation and $g_i=g_i^{eq}+g_i^{ne}$, one can obtain
\begin{subequations}\label{taylorEx-1}	
	\begin{equation}\label{gneOdt}
		g_i^{ne}=O(\Delta t),
	\end{equation}
	\begin{align}
		\sum_{l=1}^{N-1}\frac{\Delta t^{l}}{l!}D_{i}^{l}(g_i^{eq}+g_i^{ne})+\frac{\Delta t^N}{N!}D_{i}^{N}g_i^{eq}=-\Lambda_{ij}g_j^{ne}+\Delta t\left(\delta_{ij}-\frac{\Lambda_{ij}}{2}\right)F_j+O(\Delta t^{N+1}).
	\end{align}
\end{subequations}
According to Eq.\,(\ref{taylorEx-1}), we can derive the equations at different orders of $\Delta t$, 
\begin{subequations}
	\begin{equation}\label{dt1}
		D_ig_i^{eq}=-\frac{\Lambda_{ij}}{\Delta t}g_{j}^{ne}+\left(\delta_{ij}-\frac{\Lambda_{ij}}{2}\right)F_j+O(\Delta t),
	\end{equation}	
	\begin{equation}\label{dt2}
		D_i\left(g_i^{eq}+g_i^{ne}\right)+\frac{\Delta t}{2}D_i^2g_i^{eq}=-\frac{\Lambda_{ij}}{\Delta t}g_j^{ne}+\left(\delta_{ij}-\frac{\Lambda_{ij}}{2}\right)F_j+O(\Delta t^2).
	\end{equation}
\end{subequations}
From Eq.\,(\ref{dt1}), one can get
\begin{equation}\label{Ddt1}
	\frac{\Delta t}{2}D_i^2g_i^{eq}=-\frac{1}{2}D_i\Lambda_{ij}g_j^{ne}+\frac{\Delta t}{2}D_i\left(\delta_{ij}-\frac{\Lambda_{ij}}{2}\right)F_j+O(\Delta t^2).
\end{equation}
Substituting Eq.\,(\ref{Ddt1}) into Eq.\,(\ref{dt2}) yields
\begin{equation}\label{dt2-1}
	D_ig_i^{eq}+D_i\left(\delta_{ij}-\frac{\Lambda_{ij}}{2}\right)\left(g_j^{ne}+\frac{\Delta t}{2}F_j\right)=-\frac{\Lambda_{ij}}{\Delta t}g_j^{ne}+\left(\delta_{ij}-\frac{\Lambda_{ij}}{2}\right)F_j+O(\Delta t^2).
\end{equation}

To give the correct NS equations, the collision matrix $\bm{\Lambda}$ and the distribution functions $g_i^{eq}$ as well as $F_i$ should satisfy the following conditions,
\begin{subequations}\label{conditions}
	\begin{equation}
		\sum_{i}\mathbf{e}_i\Lambda_{ij}=s_0\mathbf{e}_j,\quad \sum_{i}\mathbf{c}_i\Lambda_{ij}=s_1\mathbf{c}_j,\quad
		\sum_{i}\mathbf{c}_i\mathbf{c}_i\Lambda_{ij}=s_2\mathbf{c}_j\mathbf{c}_j,
	\end{equation}
	\begin{equation}\label{sumgeq}
		\sum_{i}g_i^{eq}=\rho_0,\quad \sum_{i}\mathbf{c}_ig_i^{eq}=\rho\mathbf{u},\quad \sum_{i}\mathbf{c}_i\mathbf{c}_ig_i^{eq}=\rho\mathbf{uu}+\mathbf{m}^{\phi C}\mathbf{u}+p\mathbf{I},\quad \sum_{i}\mathbf{c}_i\mathbf{c}_i\mathbf{c}_ig_i^{eq}=c_s^2\rho\Delta\cdot\mathbf{u},
	\end{equation}
	\begin{equation}
		\sum_{i}F_i=\mathbf{u}\cdot\nabla\rho,\quad\sum_{i}\mathbf{c}_iF_i=\mathbf{F},\quad\sum_{i}\mathbf{c}_i\mathbf{c}_iF_i=\mathbf{M}_{2F},
	\end{equation}
\end{subequations}
where $\mathbf{e}_i=1$ for all $i=0,1,\cdots,q-1$. $s_0$, $s_1$ and $s_2$ are the eigenvalues of the matrix $\bm{\Lambda}$ for different eigenvectors $\left(\mathbf{e}_i\right)$, $\left(\mathbf{c}_i\right)$, and $\left(\mathbf{c}_i\mathbf{c}_i\right)$. $\Delta=\delta_{\alpha\beta}\delta_{\theta\gamma}+\delta_{\beta\theta}\delta_{\alpha\gamma}+\delta_{\alpha\theta}\delta_{\beta\gamma}$ is a four order tensor.

From Eqs.\,(\ref{comU}) and (\ref{sumgeq}), we can obtain
\begin{equation}
	\sum_{i}g_i^{ne}=-\frac{\Delta t}{2}\sum_{i}F_i=-\frac{\Delta t}{2}\mathbf{u}\cdot\nabla\rho,\quad \sum_{i}\mathbf{c}_ig_i^{ne}=-\frac{\Delta t}{2}\sum_{i}\mathbf{c}_iF_i=-\frac{\Delta t}{2}\mathbf{F}.
\end{equation}
According to above relations, we can derive the zeroth and the first-order moments of Eqs.\,(\ref{dt1}) and (\ref{dt2-1}),
\begin{subequations}
	\begin{equation}\label{NSO11}
		\partial_t\rho_0+\nabla\cdot\left(\rho\mathbf{u}\right)=\mathbf{u}\cdot\nabla\rho+O(\Delta t),
	\end{equation}
	\begin{equation}\label{NSO12}
		\partial_{t}\left(\rho\mathbf{u}\right)+\nabla\cdot\left(\rho\mathbf{uu}+\mathbf{m}^{\phi C}\mathbf{u}+p\mathbf{I}\right)=\mathbf{F}+O(\Delta t),
	\end{equation}
\end{subequations}
\begin{subequations}
	\begin{equation}\label{NSO21}
		\partial_t\rho_0+\nabla\cdot\left(\rho\mathbf{u}\right)=\mathbf{u}\cdot\nabla\rho+O(\Delta t^2),
	\end{equation}
	\begin{equation}\label{NSO22}
		\partial_{t}\left(\rho\mathbf{u}\right)+\nabla\cdot\left(\rho\mathbf{uu}+\mathbf{m}^{\phi C}\mathbf{u}+p\mathbf{I}\right)+\nabla\cdot\left(1-\frac{s_2}{2}\right)\left(\sum_{i}\mathbf{c}_i\mathbf{c}_ig_i^{ne}+\frac{\Delta t}{2}\mathbf{M}_{2F}\right)=\mathbf{F}+O(\Delta t^2),
	\end{equation}	
\end{subequations}
where the term $\sum_{i}\mathbf{c}_i\mathbf{c}_ig_i^{ne}$ can be calculated from Eq.\,(\ref{dt1}) with the help of Eq.\,(\ref{conditions}),
\begin{equation}\label{cicigi1}
	\begin{aligned}
		\sum_{i}\mathbf{c}_i\mathbf{c}_ig_i^{ne}=&-\Delta t\sum_{i}\mathbf{c}_i\mathbf{c}_i\Lambda_{ij}^{-1}\left[D_jg_j^{eq}-\left(1-\frac{s_2}{2}\right)F_j\right]+O(\Delta t^2)\\
		=&-\frac{\Delta t}{s_2}\left[\partial_{t}\left(\rho\mathbf{uu}+\mathbf{m}^{\phi C}\mathbf{u}+p\mathbf{I}\right)+\nabla\cdot \left(c_s^2\rho\Delta\cdot\mathbf{u}\right)-\left(1-\frac{s_2}{2}\right)\mathbf{M}_{2F}\right]+O(\Delta t^2)\\
		=&-\frac{\Delta t}{s_2}\left\{\partial_{t}\left(\rho\mathbf{uu}+\mathbf{m}^{\phi C}\mathbf{u}+p\mathbf{I}\right)+c_s^2\left[\mathbf{u}\nabla\rho+(\mathbf{u}\nabla\rho)^T+\mathbf{u}\cdot\nabla\rho\mathbf{I}\right]-\left(1-\frac{s_2}{2}\right)\mathbf{M}_{2F}\right\}\\
		&-\frac{c_s^2\Delta t}{s_2}\rho\left[\nabla\mathbf{u}+\left(\nabla\mathbf{u}\right)^T\right]+O(\Delta t^2).
	\end{aligned}
\end{equation}
Then we can get
\begin{equation}\label{cicigi1+}
	\begin{aligned}
		\left(1-\frac{s_2}{2}\right)\left(\sum_{i}\mathbf{c}_i\mathbf{c}_ig_i^{ne}+\frac{\Delta t}{2}\mathbf{M}_{2F}\right)=&-\left(\frac{1}{s_2}-\frac{1}{2}\right)\Delta t\left[\partial_t\left(\rho\mathbf{uu}+\mathbf{m}^{\phi C}\mathbf{u}+p\mathbf{I}\right)+c_s^2\left(\mathbf{u}\nabla\rho+(\mathbf{u}\nabla\rho)^T+\mathbf{u}\cdot\nabla\rho\mathbf{I}\right)-\mathbf{M}_{2F}\right]\\
		&-\left(\frac{1}{s_2}-\frac{1}{2}\right)\rho c_s^2\Delta t\left[\nabla\mathbf{u}+\left(\nabla\mathbf{u}\right)^T\right]+O(\Delta t^2).
	\end{aligned}
\end{equation}
If we take the following expression of $\mathbf{M}_{2F}$ and substitute Eq.\,(\ref{cicigi1+}) into Eq.\,(\ref{NSO22}),
\begin{equation}\label{M2F}
	\mathbf{M}_{2F}=\partial_t\left(\rho\mathbf{uu}+\mathbf{m}^{\phi C}\mathbf{u}+p\mathbf{I}\right)+c_s^2\left[\mathbf{u}\nabla\rho+(\mathbf{u}\nabla\rho)^T\right]+c_s^2\mathbf{u}\cdot\nabla\rho\mathbf{I},
\end{equation}
the macroscopic incompressible NS equations (\ref{NS1-1}) and (\ref{NS2-mphiC}) can be recovered at the order of $O(\Delta t^2)$ with $\nu=\left(1/s_2-1/2\right)c_s^2\Delta t$.

Additionally, from Eq.\,(\ref{cicigi1}) one can also obtain the local computing scheme for the strain rate tensor, 
\begin{equation}\label{strainRate}
	\mathbf{S}=\frac{\nabla\mathbf{u}+\left(\nabla\mathbf{u}\right)^T}{2}=-\frac{s_2}{2\rho c_s^2\Delta t}\left[\sum_{i}\left(\mathbf{c}_i\mathbf{c}_ig_i-\mathbf{c}_i\mathbf{c}_ig_i^{eq}\right)+\frac{\Delta t}{2}\mathbf{M}_{2F}\right],
\end{equation}
which is similar to that in previous work \cite{Chai2012PRE}. Then based on Eq.\,(\ref{strainRate}), we can get $\nabla\cdot\mathbf{u}=\mathbf{tr}\left(\mathbf{S}\right)$ with $\mathbf{tr}\left(\mathbf{S}\right)$ being the trace of matrix $\mathbf{S}$.

\subsection{The computation of pressure}\label{computingP}
Now let us focus on the computation of pressure. From Eq.\,(\ref{dt1}) one can obtain
\begin{equation}\label{gne}
	g_i^{ne}=-\Delta t\Lambda_{ij}^{-1}\left[D_jg_j^{eq}-\left(\delta_{jk}-\frac{\Lambda_{jk}}{2}\right)F_k\right]+O(\Delta t^2).
\end{equation}
Considering the distribution function at the zeroth direction ($g_0=g_0^{eq}+g_0^{ne}$) with the following $g_0^{eq}$ and $g_0^{ne}$,
\begin{subequations}
	\begin{equation}
		g_0^{eq}=\frac{\omega_0-1}{c_s^2}p+\rho_0+s_0\left(\phi,\mathbf{u}\right),
	\end{equation}
	\begin{equation}
		\begin{aligned}
			g_0^{ne}=&-\Delta t\partial_{t}\left(\Lambda_{0k}^{-1}g_k^{eq}\right)+\Delta t\left(\Lambda_{0k}^{-1}F_k-\frac{F_0}{2}\right)+O(\Delta t^2)\\
			=&H\Delta t\mathbf{u}\cdot\nabla\rho+Jd\Delta t\partial_tp+K\Delta t\partial_t\left(\rho\mathbf{u}\cdot\mathbf{u}+\mathbf{m}^{\phi C}\cdot\mathbf{u}\right)+O(\Delta t^2),
		\end{aligned}
	\end{equation}
\end{subequations}
we have
\begin{equation}
	\begin{aligned}
		\frac{1-\omega_0}{c_s^2}p=&\rho_0-\left(g_0- g_0^{ne}\right)+s_0\left(\phi,\mathbf{u}\right)\\
		=&\rho_0-\sum_{i}g_i^{eq}+\frac{\Delta t}{2}\sum_{i}F_i+\sum_{i\neq0}g_i+s_0\left(\phi,\mathbf{u}\right)+H\Delta t\mathbf{u}\cdot\nabla\rho+Jd\Delta t\partial_tp+K\Delta t\partial_t\left(\rho\mathbf{u}\cdot\mathbf{u}+\mathbf{m}^{\phi C}\cdot\mathbf{u}\right)+O(\Delta t^2)\\
		=&\sum_{i\neq0}g_i+\left(\frac{1}{2}+H\right)\Delta t\mathbf{u}\cdot\nabla\rho+s_0\left(\phi,\mathbf{u}\right)+Jd\Delta t\partial_tp+K\Delta t\partial_t\left(\rho\mathbf{u}\cdot\mathbf{u}+\mathbf{m}^{\phi C}\cdot\mathbf{u}\right)+O(\Delta t^2).
	\end{aligned}
\end{equation}

Ignoring the truncation error term $O(\Delta t^2)$, one can obtain the computational scheme for pressure,
\begin{equation}\label{computP}
	p=\frac{c_s^2}{1-\omega_0}\left[\sum_{i\neq0}g_i+\left(\frac{1}{2}+H\right)\Delta t\mathbf{u}\cdot\nabla\rho+s_0\left(\phi,\mathbf{u}\right)+Jd\Delta t\partial_tp+K\Delta t\partial_t\left(\rho\mathbf{u}\cdot\mathbf{u}+\mathbf{m}^{\phi C}\cdot\mathbf{u}\right)\right].
\end{equation}

\begin{remark}\label{3Scheme}
	In the implementation of present LB model, there are three ways to deal with the term $\partial_t\left(\rho\mathbf{uu}+p\mathbf{I}\right)$ in Eq.\,(\ref{M2F}). In the first way, we can directly ignore this term as $O(Ma^2)$ with $Ma$ being the Mach number, in this case, the terms $\partial_tp$ and $\partial_t\left(\rho\mathbf{u}\cdot\mathbf{u}\right)$ in Eq.\,(\ref{computP}) are also neglected. In the second way, like the previous work \cite{Wang2019Capillarity}, we can simplify this term as $\mathbf{uF}+\mathbf{Fu}+O(Ma\Delta t+Ma^2)$, $\partial_tp$ is neglected and $\partial_t\left(\rho\mathbf{u}\cdot\mathbf{u}\right)$ can be derived as $2\mathbf{F}\cdot\mathbf{u}+O(Ma\Delta t+Ma^2)$ (some details are shown in \ref{Prhouu}). In the third way, this term is completely remained in $\mathbf{M}_{2F}$ such that the computational scheme of pressure need to be modified due to the existence of $\partial_tp$. However, our preliminary results show that the last way is unstable, although there is no low Mach number assumption. In this work, we will apply the second way in the numerical simulations.      
\end{remark}
\begin{remark}
	If we set $\bm{\Lambda}=\mathbf{M}^{-1}\mathbf{S}_g\mathbf{M}$, $\mathbf{S}_g=\mathbf{diag}(s_g^0, s_g^1, s_g^2, \cdots,s_g^{q-1})$ is a diagonal relaxation matrix, $\mathbf{M}$ is the transformation matrix composed of the orthogonal or nonorthogonal eigenvectors, the evolution equation can be written as
	\begin{equation}
		\begin{aligned}
			g_i(\mathbf{x}+\mathbf{c}_i\Delta t,t+\Delta t)=&g_i(\mathbf{x},t)-\left(\mathbf{M}^{-1}\mathbf{S}_g\mathbf{M}\right)_{ij}\left[g_j(\mathbf{x},t)-g_j^{eq}(\mathbf{x},t)\right]+\Delta t\left[\mathbf{M}^{-1}\left(\mathbf{I}-\mathbf{S}_g/2\right)\mathbf{M}\right]_{ij}F_j(\mathbf{x},t)\\
			=&\mathbf{M}_{ij}^{-1}\left[\left(\mathbf{I}-\mathbf{S}_g\right)\mathbf{m}_g+\mathbf{S}_g\mathbf{m}_g^{eq}+\Delta t \left(\mathbf{I}-\mathbf{S}_g/2\right)\mathbf{m}_F\right]_j,	
		\end{aligned}
	\end{equation}
	where $\mathbf{m}_g=\left(\mathbf{M}_{ij}g_j\right)$, $\mathbf{m}^{eq}=\left(\mathbf{M}_{ij}g_j^{eq}\right)$ and $\mathbf{m}_F=\left(\mathbf{M}_{ij}F_j\right)$.   
\end{remark}

\section{Numerical results and discussion}\label{numerical}
In this section, several two-dimensional benchmark problems, including the deformation of a square droplet, the layered Poiseuille flow, a droplet spreading on an ideal wall, the RTI and the dam break, are considered to test the present LB model. In the following, to give a comparison with some available works, the CH equation with $\phi_A=1$ and $\phi_B=-1$ is adopted for problems in the subsections \ref{square} and \ref{RTI}, while AC equation with $\phi_A=1$ and $\phi_B=0$ is applied to the problems in subsections \ref{Poiseuille}, \ref{DroSp} and \ref{Dambreak}. Here the original LB model \cite{Fakhari2010IJNMF,Liang2018PRE} is considered as the one with the flux $\mathbf{m}=\rho\mathbf{u}$, and the corrected LB model denotes the one with the flux $\mathbf{m}=\rho\mathbf{u}$ and the additional interface force $\mathbf{F}_{a}=\frac{d\rho}{d\phi}\mathbf{u}\nabla\cdot M_{\phi}\mathbf{D}\left(\phi\right)$ \cite{Li2012PRE, Ren2016PRE}. In the following simulations, the D2Q9 lattice model (the transformation matrix $\mathbf{M}$ and the moments of distribution functions are shown in \ref{D2Q9}) is adopted for both phase-field and NS equations, and the half-way bounce-back scheme \cite{Ladd1994JFM1,Ladd1994JFM2} is applied for the no-flux and no-slip velocity boundary conditions. The relaxation parameters corresponding to mobility and viscosity are given by $s_f^3=s_f^5=1/\left(M_{\phi}/\eta c_s^2\Delta t+0.5\right)$, $s_g^7=s_g^8=1/\left(\nu/c_s^2\Delta t+0.5\right)$, while the others are set to be 1 if not specified. 

\subsection{The deformation of a square droplet}\label{square}
The deformation of a square droplet is a simple two-phase problem, and with the time increases, the square droplet would deform into a circle one under the action of the surface tension. In this part, we will consider this problem with a large density ratio $\rho_A/\rho_B=1000:1$ to show the incompressibility and the consistency of mass conservation of the LB model. Initially, a square droplet (phase A) with the length $D=\sqrt{\pi R^2+\left(4-\pi\right)W^2/4}$ is located at the center of the square domain $[-1,1]\times[-1,1]$, and is surrounded by the fluid B, $R$ is the radius of the circular droplet formed at the final state. The  periodic boundary condition is applied at all boundaries, and the initial distribution of order parameter is set as
\begin{equation}
	\phi\left(x,y\right)=\begin{cases}
		\frac{\phi_A+\phi_B}{2}+\frac{\phi_A-\phi_B}{2}\tanh\frac{2\left[0.5W-\sqrt{\left(x-0.5(W-D)\right)^2+\left(y-0.5(W-D)\right)^2}\right]}{W},\ x\leq 0.5(W-D), y\leq 0.5(W-D),\\	
		\frac{\phi_A+\phi_B}{2}+\frac{\phi_A-\phi_B}{2}\tanh\frac{2\left[0.5W-\sqrt{\left(x-0.5(D-W)\right)^2+\left(y-0.5(W-D)\right)^2}\right]}{W},\ x\geq 0.5(D-W), y\leq 0.5(W-D),\\	
		\frac{\phi_A+\phi_B}{2}+\frac{\phi_A-\phi_B}{2}\tanh\frac{2\left[0.5W-\sqrt{\left(x-0.5(W-D)\right)^2+\left(y-0.5(D-W)\right)^2}\right]}{W},\ x\leq 0.5(W-D), y\geq 0.5(D-W),\\		
		\frac{\phi_A+\phi_B}{2}+\frac{\phi_A-\phi_B}{2}\tanh\frac{2\left[0.5W-\sqrt{\left(x-0.5(D-W)\right)^2+\left(y-0.5(D-W)\right)^2}\right]}{W},\ x\geq 0.5(D-W), y\geq 0.5(D-W),\\
		\frac{\phi_A+\phi_B}{2}+\frac{\phi_A-\phi_B}{2}\tanh\frac{2\left(0.5D+y\right)}{W},\ 0.5(W-D)<x<0.5(D-W), y\leq 0.5(W-D),\\
		\frac{\phi_A+\phi_B}{2}+\frac{\phi_A-\phi_B}{2}\tanh\frac{2\left(0.5D-y\right)}{W},\ 0.5(W-D)<x<0.5(D-W), y\geq 0.5(D-W),\\
		\frac{\phi_A+\phi_B}{2}+\frac{\phi_A-\phi_B}{2}\tanh\frac{2\left(0.5D+x\right)}{W},\ x\leq 0.5(W-D), 0.5(W-D)<y<0.5(D-W),\\
		\frac{\phi_A+\phi_B}{2}+\frac{\phi_A-\phi_B}{2}\tanh\frac{2\left(0.5D-x\right)}{W},\ x\geq 0.5(D-W), 0.5(W-D)<y<0.5(D-W),\\	
		\phi_A,\ \text{otherwise}.
	\end{cases}
\end{equation}
In our simulations, some physical parameters are given by $\nu_A=\nu_B=0.1$, $M_{\phi}=0.1$, $\sigma=0.001$, $R=0.5$, $W=4\Delta x=0.04$, and the particle speed is $c=\Delta x/\Delta t=100$. 

To quantify the incompressibility and the consistency of mass conservation, the root-mean-square (rms) values of the velocity divergence and mass source are used,
\begin{equation}
	D_u^{rms}=\sqrt{\langle\left(\nabla\cdot\mathbf{u}\right)^2\rangle}=\left[\frac{\sum_{ij}\left(\nabla\cdot\mathbf{u}\right)_{ij}^2}{NxNy}\right]^{1/2},
\end{equation}
\begin{equation}
	S_m^{rms}=\sqrt{\langle\left(\partial_t\rho+\nabla\cdot\mathbf{m}^{CC}\right)^2\rangle}=\left[\frac{\sum_{ij}\left(\partial_t\rho+\nabla\cdot\mathbf{m}^{CC}\right)_{ij}^2}{NxNy}\right]^{1/2},
\end{equation}
where $(\cdot)_{ij}$ represents the variable at position $(i\Delta x, j\Delta x)$, the velocity divergence is the trace of the strain rate tensor $\mathbf{S}$, and can be calculated by Eq.\,(\ref{strainRate}). The term $\nabla\cdot\mathbf{m}^{\phi C}$ in $\nabla\cdot\mathbf{m}^{CC}$ is computed by the second-order isotropic difference scheme given by Eq.\,(\ref{central1}).

\begin{figure}
	\centering
	\subfigure[]{
		\begin{minipage}{0.49\linewidth}
				\centering
				\includegraphics[width=2.5in]{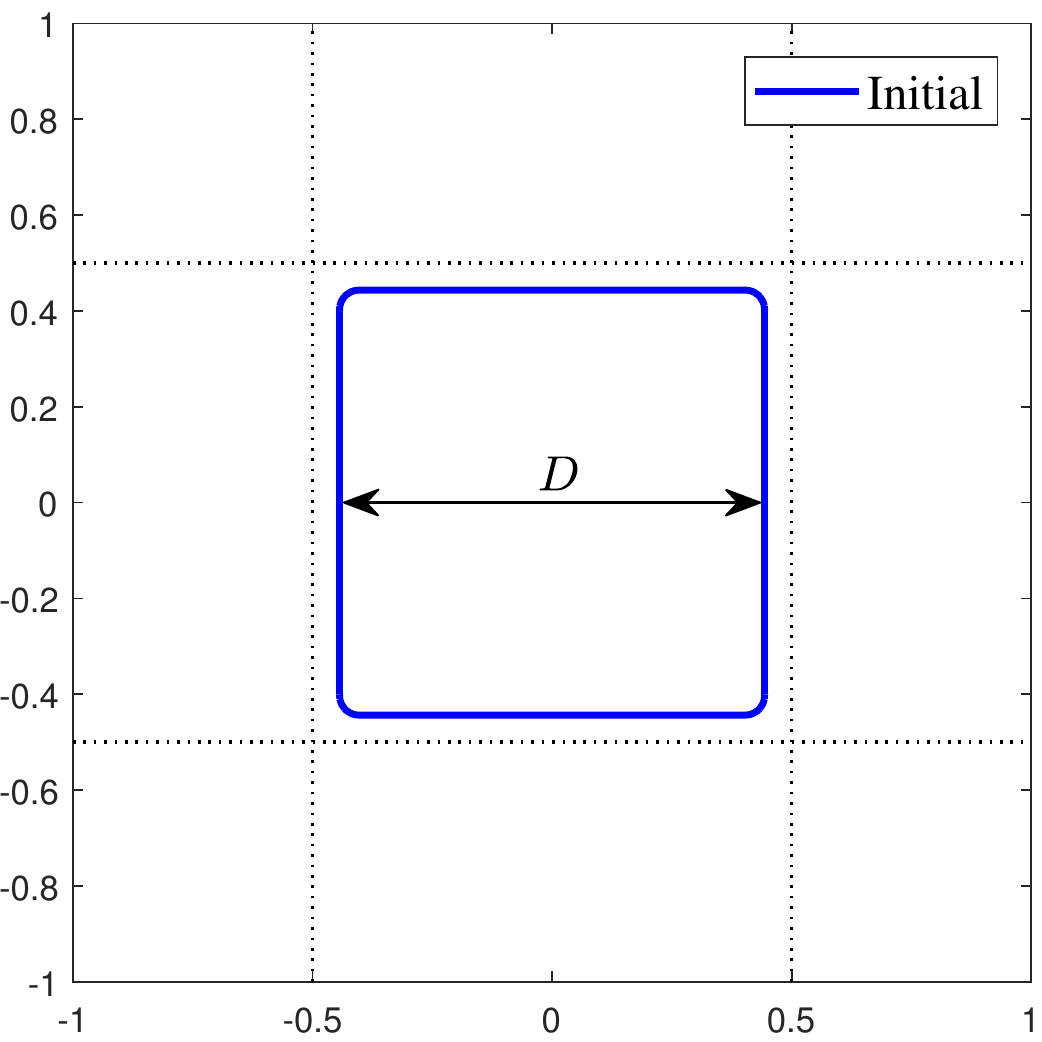}
				\label{fig-square0}
\end{minipage}}	
	\subfigure[]{
		\begin{minipage}{0.49\linewidth}
			\centering
			\includegraphics[width=2.5in]{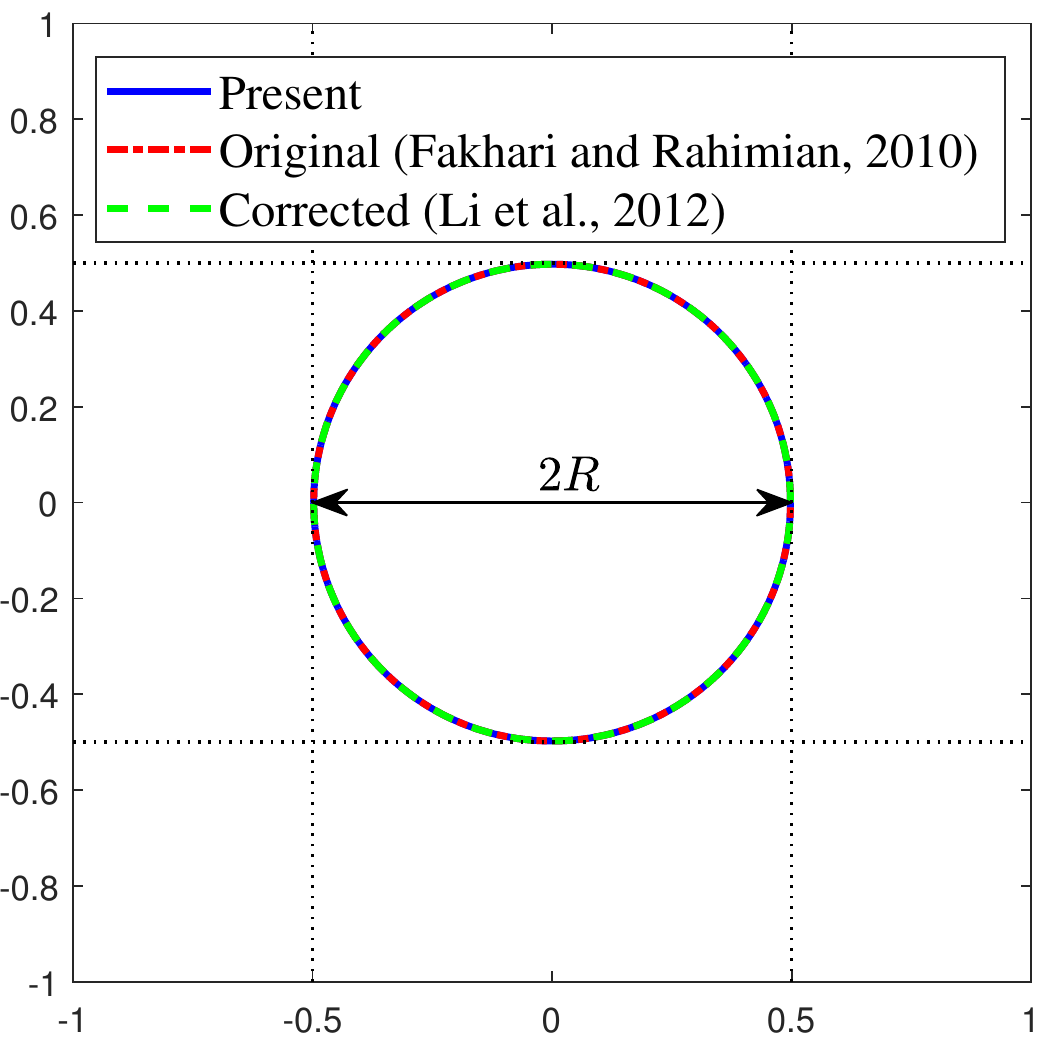}
			\label{fig-square1}
	\end{minipage}}	
	\caption{The shapes of the droplet before and after deformation [(a) The initial state, (b) The equilibrium state].}
	\label{fig-squareShape}
\end{figure}

\begin{figure}
	\centering
	\subfigure[]{
		\begin{minipage}{0.49\linewidth}
			\centering
			\includegraphics[width=3.0in]{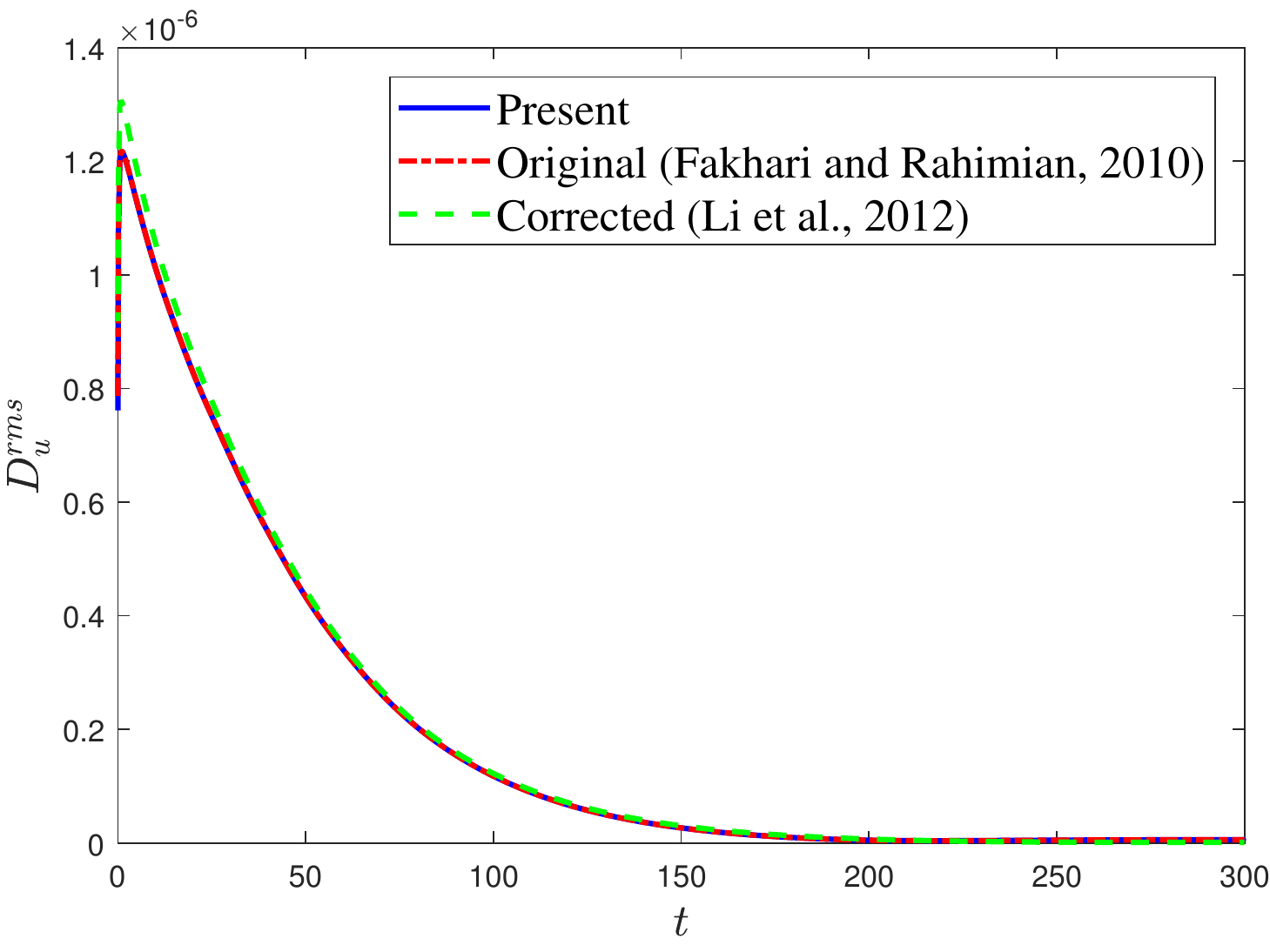}
			\label{fig-Du}
	\end{minipage}}	
	\subfigure[]{
		\begin{minipage}{0.49\linewidth}
			\centering
			\includegraphics[width=3.0in]{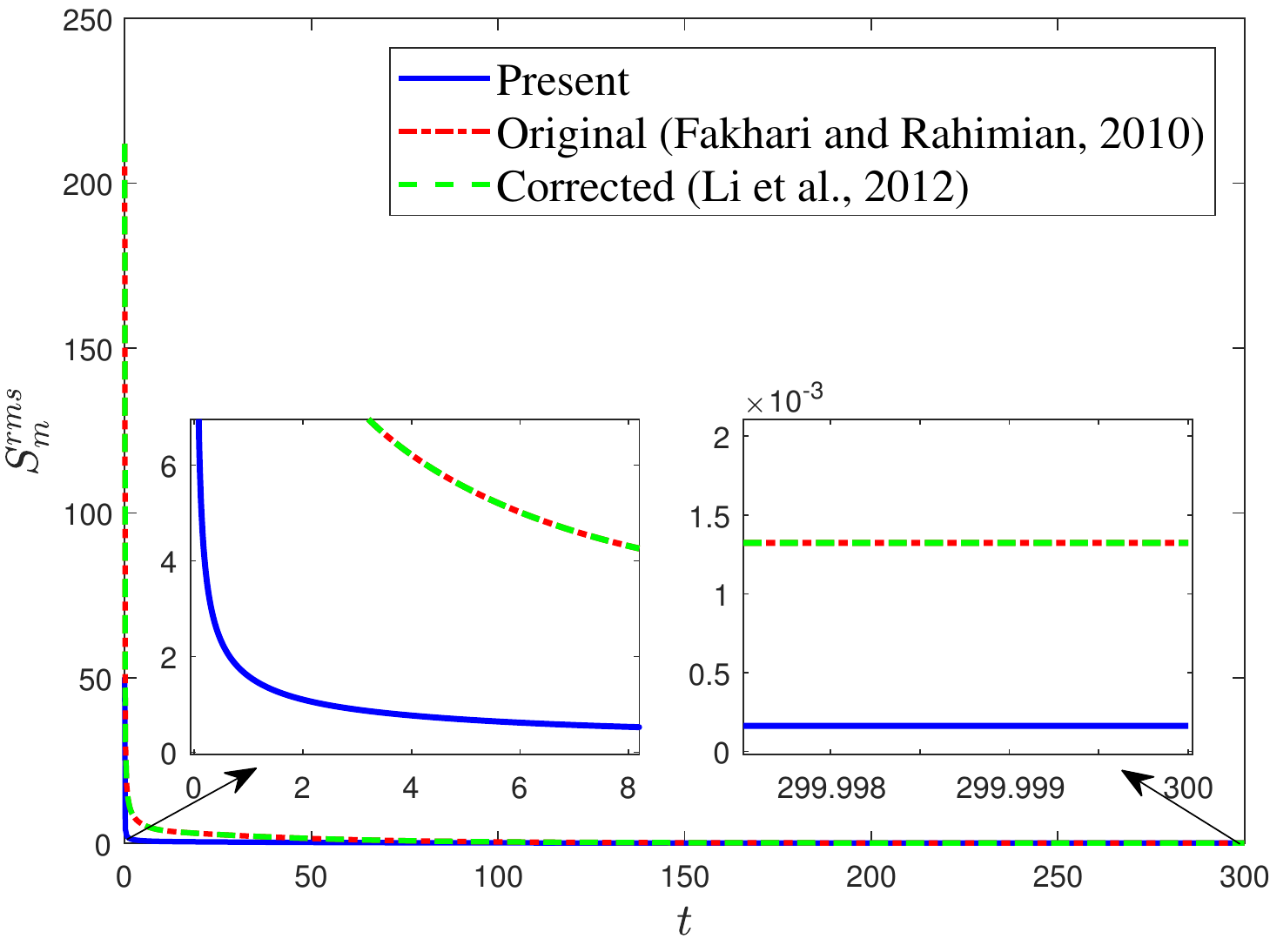}
			\label{figSm}
	\end{minipage}}	
	\caption{Evolution of the statistical variables [(a) $D_u^{rms}$, (b) $S_m^{rms}$] in time.}
	\label{fig-square-D}
\end{figure}

The droplet shapes before and after deformation are shown in Fig.\,\ref{fig-squareShape}. From this figure, one can find that the numerical results of the present LB model agree well with those of the original and corrected LB models, and the radius of the droplet equals to 0.5 at the final equilibrium state, which is also in good agreement with the specified value. Additionally, as shown in Fig.\,\ref{fig-square-D}, the changes of $D_u^{rms}$ based on three LB models are similar, while the value of $S_m^{rms}$ obtained by present LB model is much smaller than those of the original and corrected models, which can be seen clearly from the Table \ref{table-Square}. This indicates that present LB model is better in preserving the consistency of mass conservation.
\begin{table}
	\centering
	\caption{The values of $D_u^{rms}$ and $S_{m}^{rms}$ of the deformation of a square droplet at steady state ($t=300$).}
	\begin{tabular}{ccccccc}
		\toprule
		&& Original && Corrected && Present \\
		\midrule
		$D_u^{rms}$ && $5.9254\times10^{-9}$ && $1.1827\times10^{-9}$ && $5.9254\times10^{-9}$ \\
		$S_{m}^{rms}$  && $1.3232\times10^{-3}$ && $1.3232\times10^{-3}$ && $1.6241\times10^{-4}$\\
		\bottomrule
	\end{tabular}
	\label{table-Square}
\end{table}

\subsection{The layered Poiseuille flow}\label{Poiseuille}
We now consider the layered Poiseuille flow, which is a two-phase channel flow driven by a constant force $\mathbf{G}=(G_x,0)$. In the channel, the fluid A with density $\rho_A$ is placed in the lower region of $-h\leq y\leq 0$, and the upper half region $0<y\leq h$ is filled with fluid B. The periodic boundary condition is applied to the left and right boundaries, while the no-slip velocity boundary condition is imposed on the top and bottom boundaries. For this problem, the initial distribution of order parameter is approximated by
\begin{equation}
	\phi\left(x,y\right)=\frac{\phi_A+\phi_B}{2}+\frac{\phi_A-\phi_B}{2}\tanh\frac{-2y}{W},
\end{equation} 
in this case, the linear dynamic viscosity given by Eq.\,(\ref{linearMu}) can be expressed as
\begin{equation}
	\mu\left(x,y\right)=\frac{\mu_A+\mu_B}{2}+\frac{\mu_A-\mu_B}{2}\tanh\frac{-2y}{W}.
\end{equation} 
We apply above linear dynamic viscosity such that the smooth analytical solution of the velocity in the horizontal direction can be derived,
\begin{equation}
	U_e(y)=-\frac{U_c}{AD/B-C}\int_{-h}^{y}\frac{y'}{\mu}dy'+\frac{U_c}{D-BC/A}\int_{-h}^{y}\frac{1}{\mu}dy',
\end{equation}
where $U_c$ represents the steady horizontal velocity at the center line, $A=\int_{-h}^{h}\frac{y}{\mu}dy$, $B=\int_{-h}^{h}\frac{1}{\mu}dy$, $C=\int_{-h}^{0}\frac{y}{\mu}dy$, and $D=\int_{-h}^{0}\frac{1}{\mu}dy$.  

To quantitatively evaluate the accuracy of the present LB model, the following relative error is adopted,
\begin{equation}
	Err_u=\frac{\sum_j\lvert U_x(y_j,t_n)-U_e(y_j)\rvert}{\sum_j\lvert U_e(y_j)\rvert},
\end{equation}
where $y_j=j\Delta x$, $t_n=n\Delta t$, $U_x$ denotes the numerical solution.

\begin{figure}
	\centering
	\subfigure[]{
		\begin{minipage}{0.49\linewidth}
			\centering
			\includegraphics[width=3.0in]{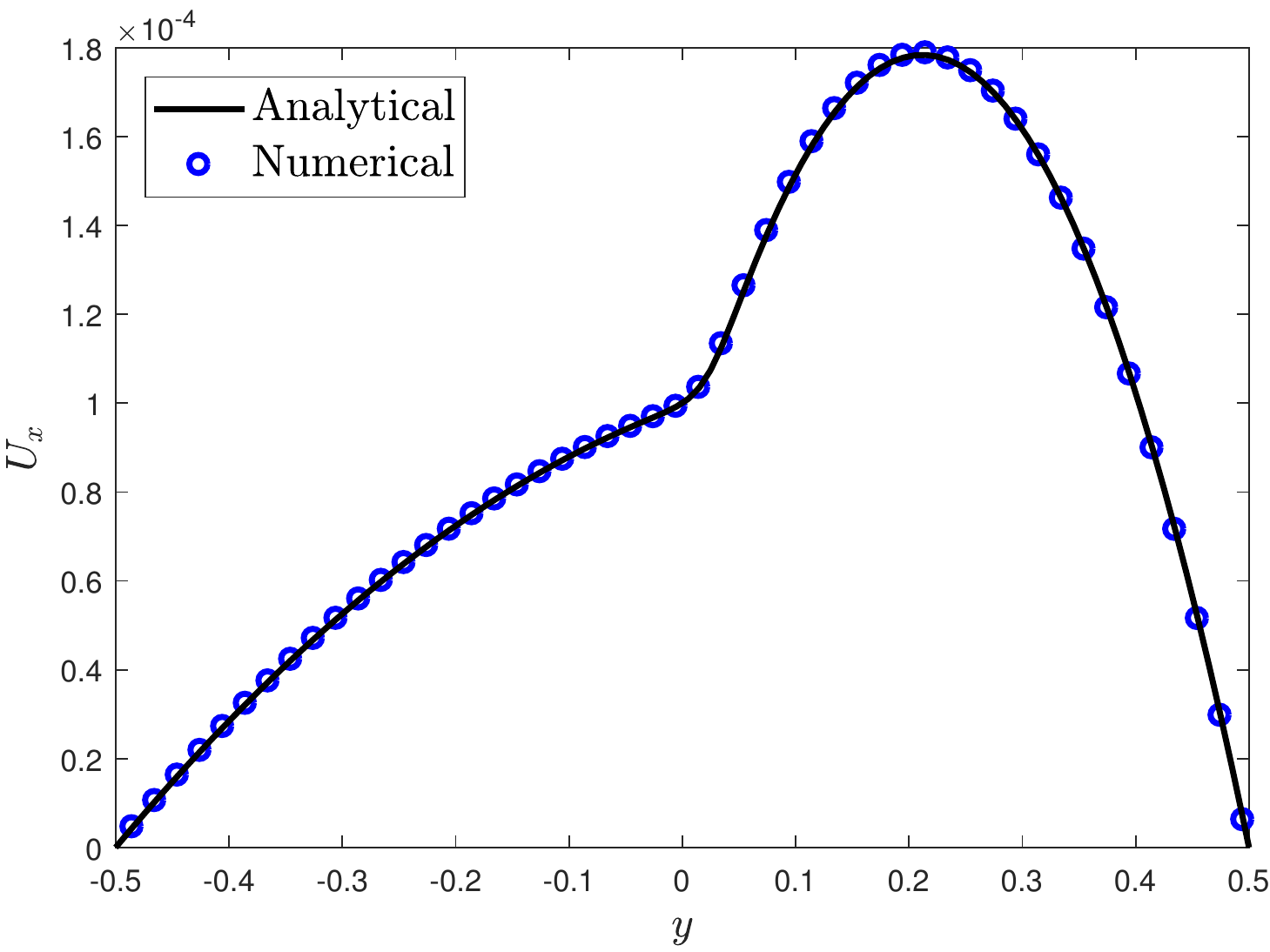}
			\label{fig-rho10}
	\end{minipage}}	
	\subfigure[]{
		\begin{minipage}{0.49\linewidth}
			\centering
			\includegraphics[width=3.0in]{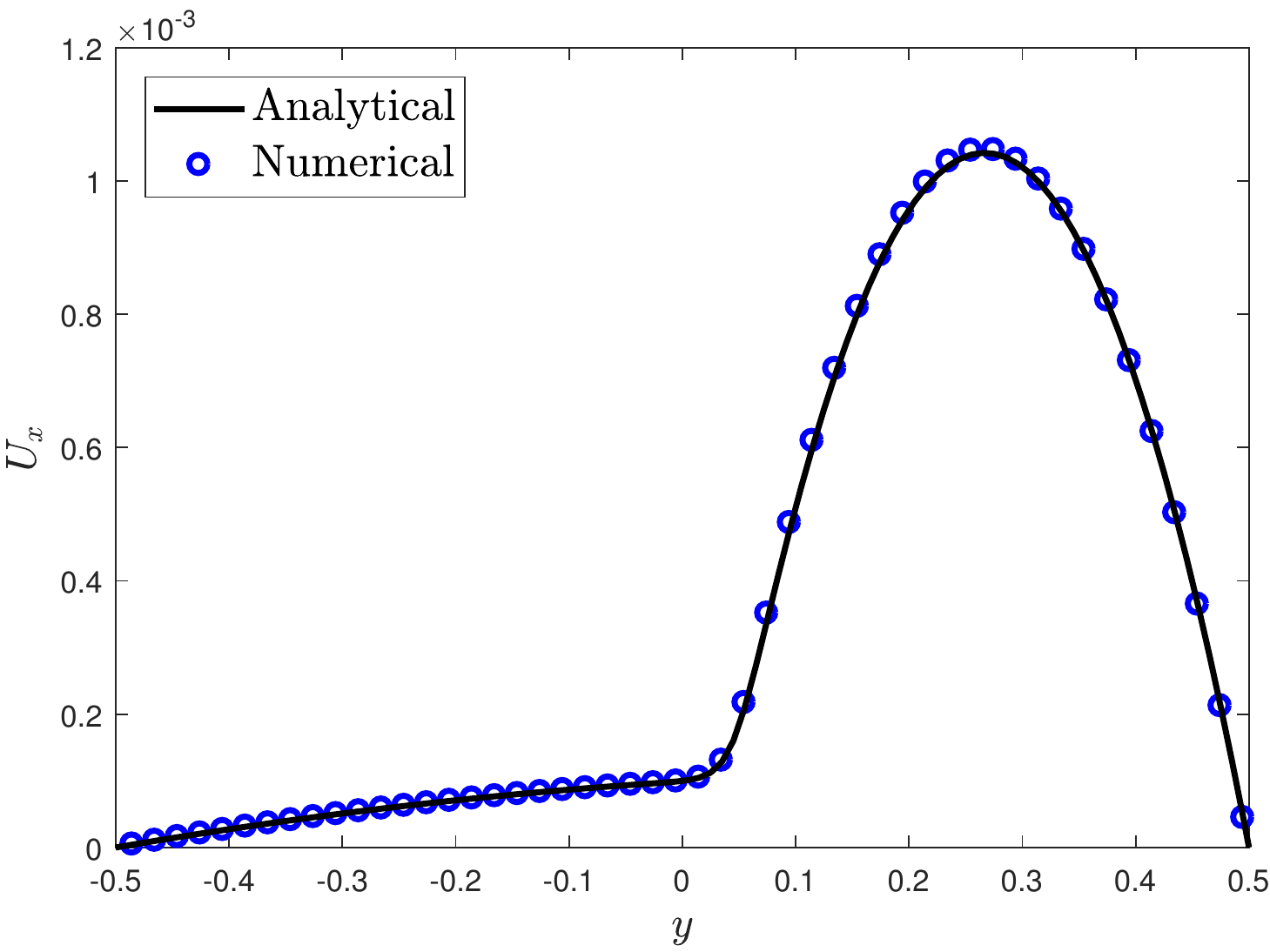}
			\label{fig-rho100}
	\end{minipage}}	
	
	\subfigure[]{
		\begin{minipage}{0.49\linewidth}
			\centering
			\includegraphics[width=3.0in]{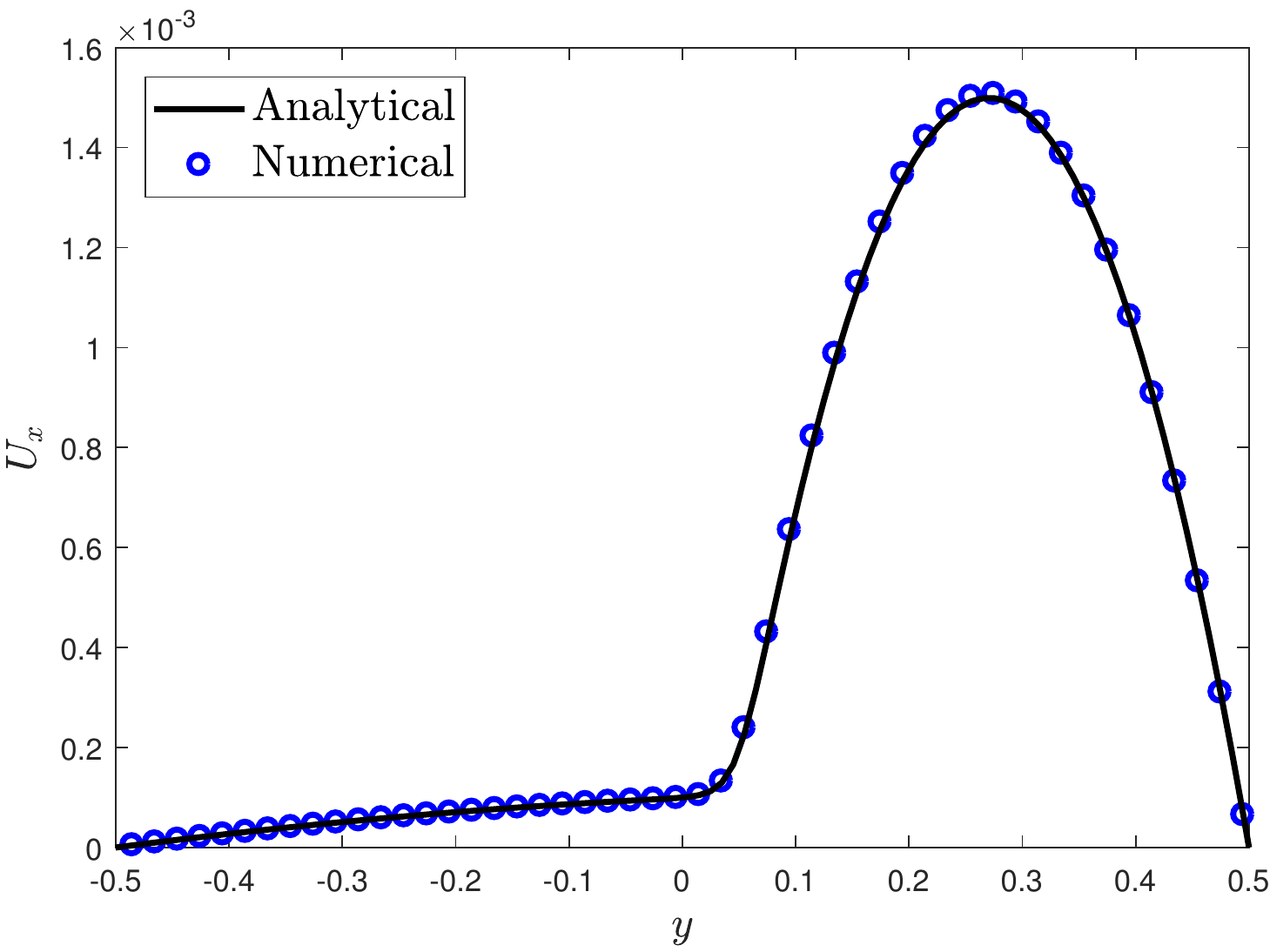}
			\label{fig-rho150}
	\end{minipage}}	
	\subfigure[]{
		\begin{minipage}{0.49\linewidth}
			\centering
			\includegraphics[width=3.0in]{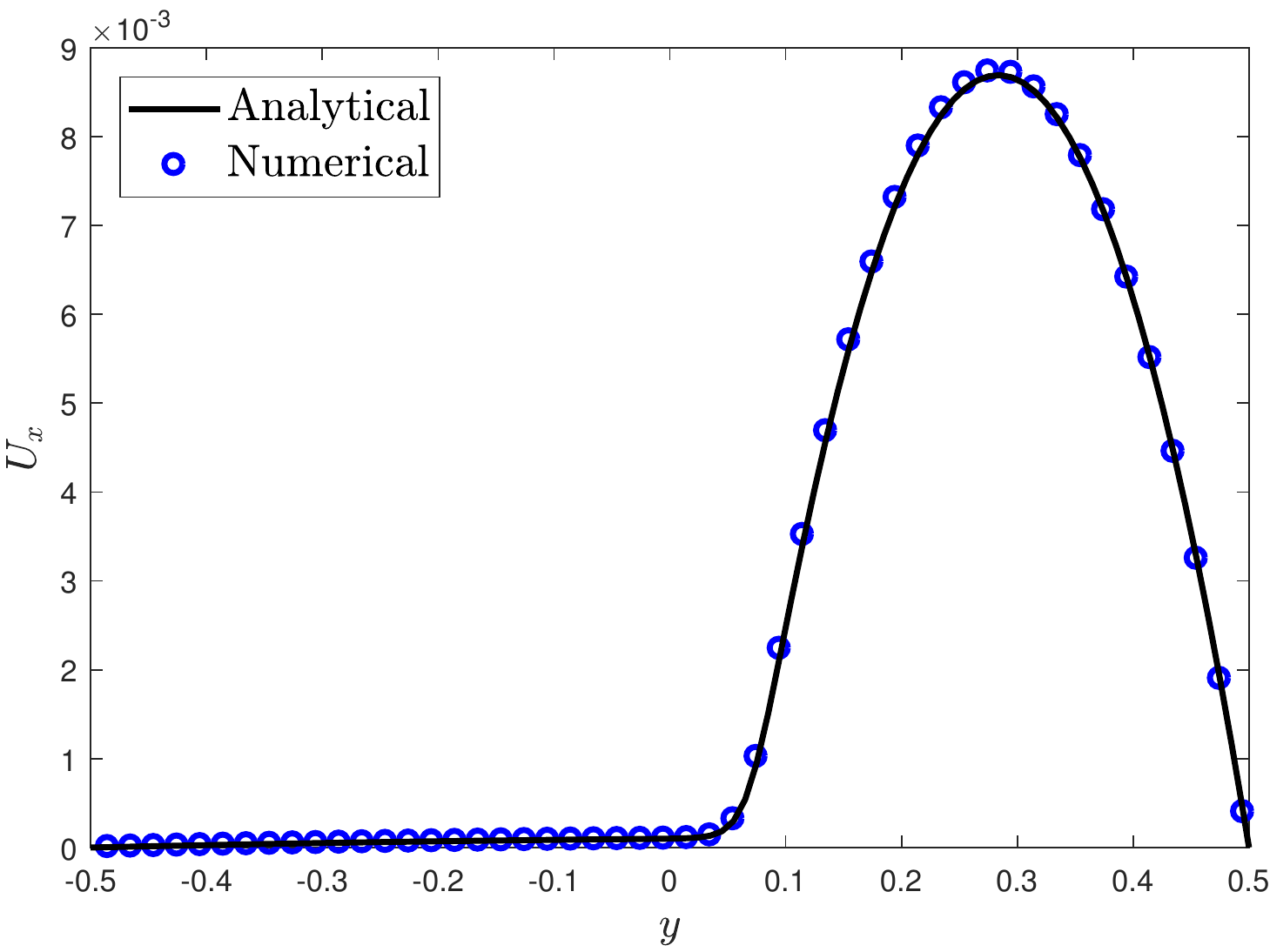}
			\label{fig-rho1000}
	\end{minipage}}	
	\caption{A comparison of the numerical and analytical solutions of the horizontal velocity at different density ratios [(a) $\rho_A/\rho_B=10:1$, (b) $\rho_A/\rho_B=100:1$, (c) $\rho_A/\rho_B=150:1$, (d) $\rho_A/\rho_B=1000:1$].}
	\label{fig-PoiRho}
\end{figure}

In our simulations, the grid is set as $Nx\times Ny=10\times100$, $U_c$ is fixed to be $10^{-4}$, the other parameters are given by $h=0.5$, $W=4\Delta x$, $c=100$, $\sigma=0.001$, $\nu_A=\nu_B=0.1$ and $M_{\phi}=0.1$. We presented the profiles of the horizontal velocity $U_x$ at different density ratios in Fig.\,\ref{fig-PoiRho}, and found that the numerical results of present LB model are in agreement with the analytical solutions. In addition, the relative errors of the horizontal velocity $U_x$ are also listed in Table \ref{table-PoiErr}, and it is found that these errors are smaller than those in some previous works where a step function of viscosity is used \cite{Fakhari2017JCP,Ren2016PRE,Liang2018PRE}. 

\begin{table}
	\centering
	\caption{The relative errors of the horizontal velocity $U_x$ of the layered Poiseuille flow.}
	\begin{tabular}{ccccccccc}
		\toprule
		Density ratio && $10:1$  &&  $100:1$  &&  $150:1$  &&  $1000:1$ \\
		\midrule
		Relative error   && $1.538\times10^{-3}$ &&  $6.569\times10^{-3}$ && $7.164\times10^{-3}$ && $8.366\times10^{-3}$  \\
		\bottomrule
	\end{tabular}
	\label{table-PoiErr}
\end{table}
\begin{figure}
	\centering
	\subfigure[]{
		\begin{minipage}{0.3\linewidth}
			\centering
			\includegraphics[width=1.8in]{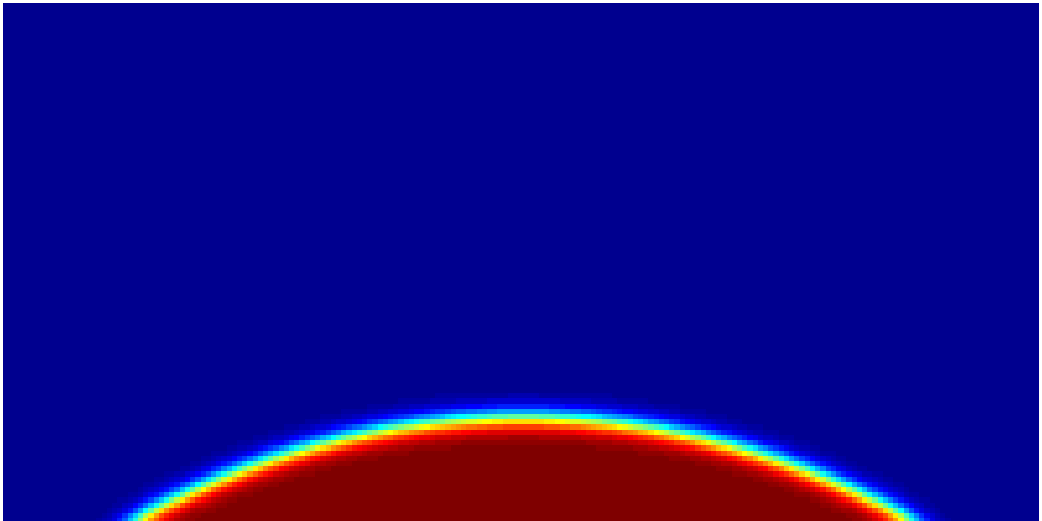}
			\label{fig-Sp30}
	\end{minipage}}	
	\subfigure[]{
		\begin{minipage}{0.3\linewidth}
			\centering
			\includegraphics[width=1.8in]{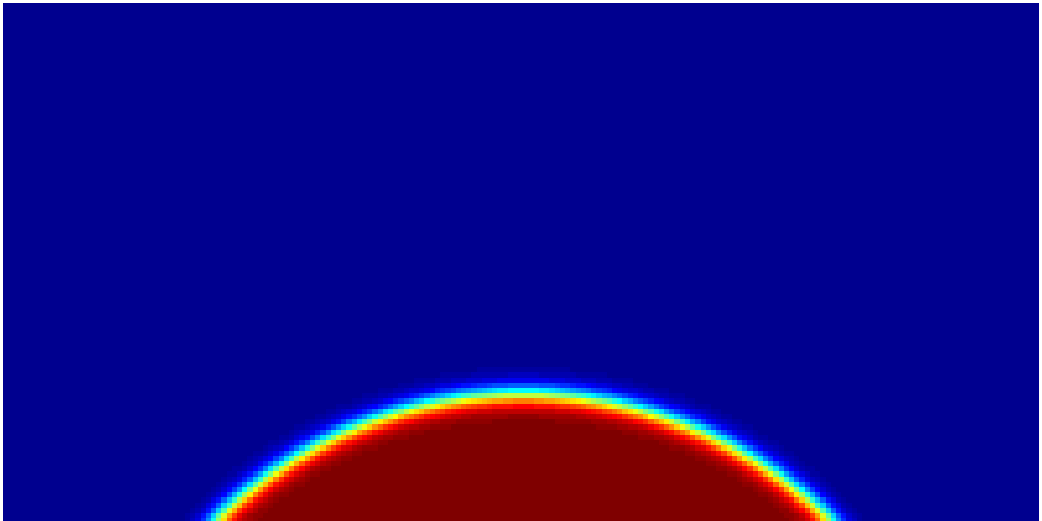}
			\label{fig-Sp45}
	\end{minipage}}		
	\subfigure[]{
		\begin{minipage}{0.3\linewidth}
			\centering
			\includegraphics[width=1.8in]{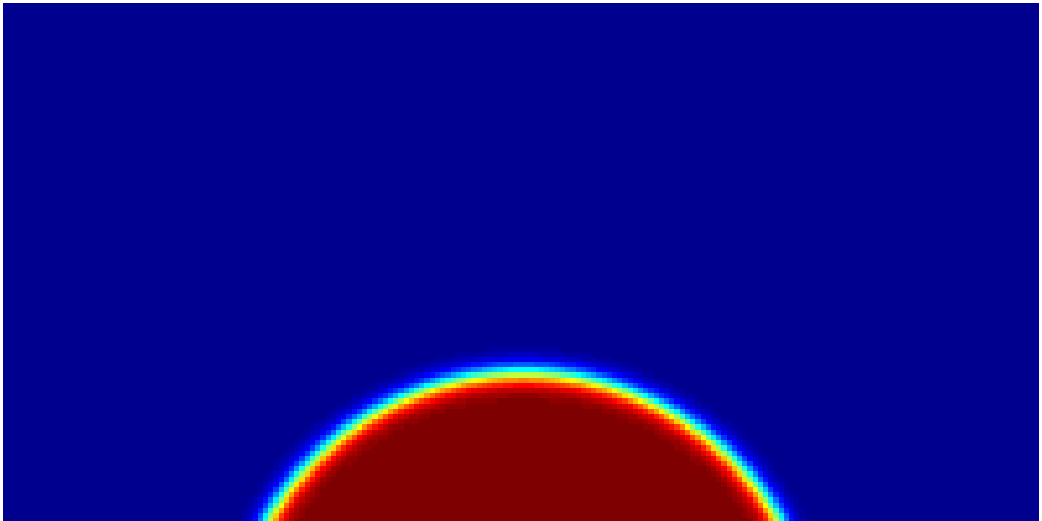}
			\label{fig-Sp60}
	\end{minipage}}	
	
	\subfigure[]{
		\begin{minipage}{0.3\linewidth}
			\centering
			\includegraphics[width=1.8in]{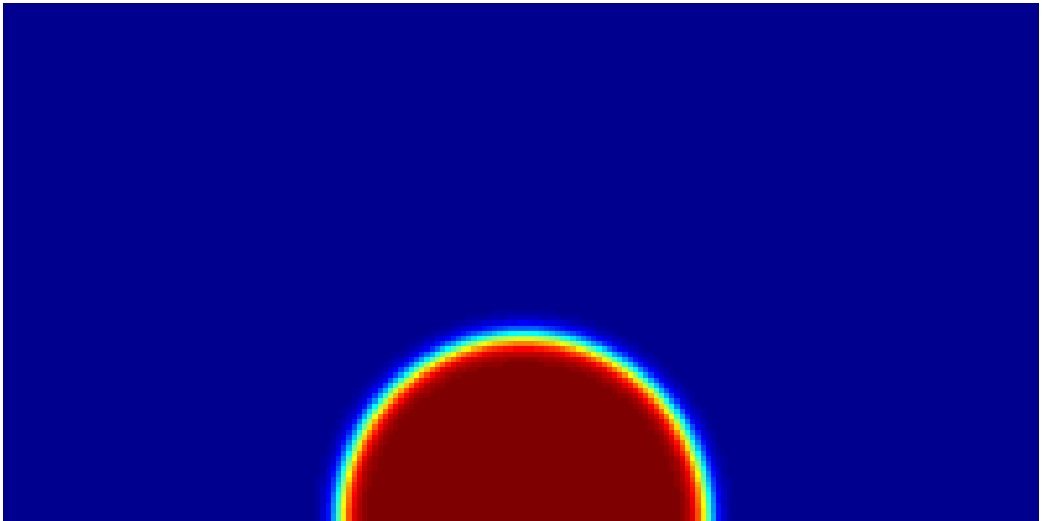}
			\label{fig-Sp90}
	\end{minipage}}	
	\subfigure[]{
		\begin{minipage}{0.3\linewidth}
			\centering
			\includegraphics[width=1.8in]{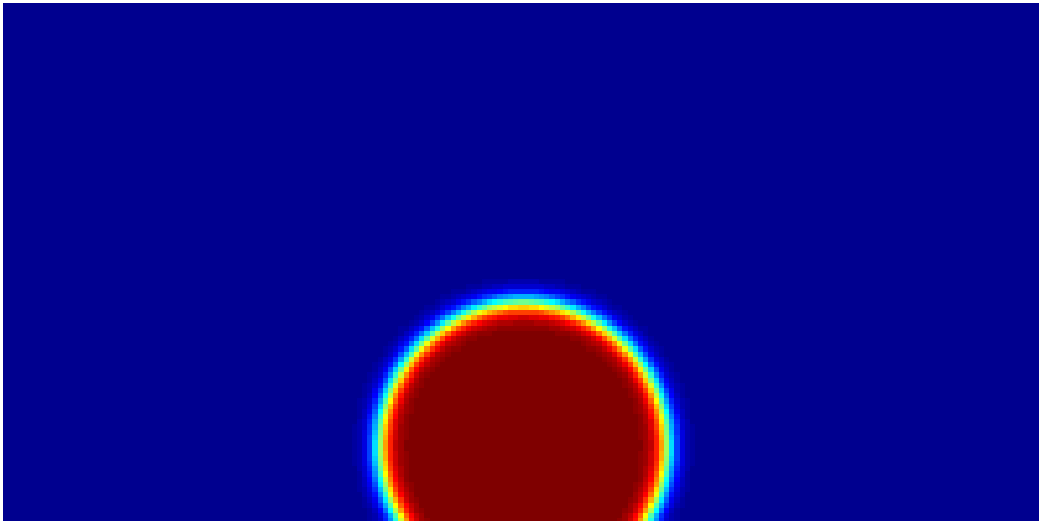}
			\label{fig-Sp120}
	\end{minipage}}	
	\subfigure[]{
		\begin{minipage}{0.3\linewidth}
			\centering
			\includegraphics[width=1.8in]{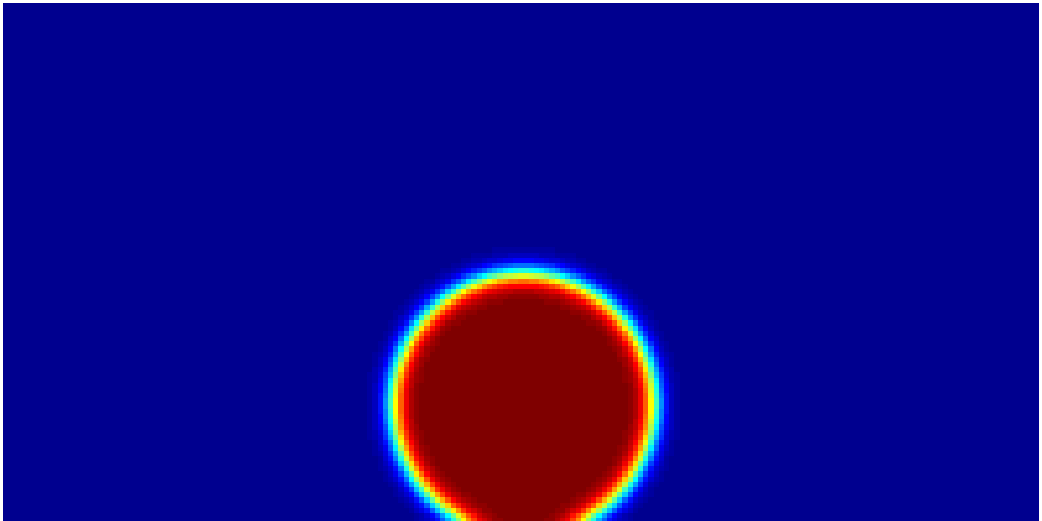}
			\label{fig-Sp150}
	\end{minipage}}	
	\caption{The predicted equilibrium shapes of the droplet under different prescribed contact angles [(a) $\theta=30\degree$, (b) $\theta=45\degree$, (c) $\theta=60\degree$, (d) $\theta=90\degree$, (e) $\theta=120\degree$, (f) $\theta=150\degree$].}
	\label{fig-Sp}
\end{figure}

\subsection{A droplet spreading on an ideal wall}\label{DroSp}
The third problem we considered is a droplet spreading on an ideal wall, which would be used to test the capacity of the present LB model in predicting the contact angle. In this problem, the density ratio ($\rho_A/\rho_B$) and dynamic viscosity ratio ($\mu_A/\mu_B$) are fixed to be 1000 and 100, which are close to the values of the realistic water-air system at room temperature and normal atmospheric pressure. The simulations are performed in the computational domain $[-100,100]\times[0,100]$, in which a semicircular droplet with the radius $R_0=35$ is initially deposited on the bottom wall. The periodic boundary condition is used in the horizontal direction, while the wetting and no-flux boundary conditions are imposed at the bottom and top boundaries. The distribution profile of the order parameter is initialized by
\begin{equation}
	\phi\left(x,y\right)=\frac{\phi_A+\phi_B}{2}+\frac{\phi_A-\phi_B}{2}\tanh\frac{2\left(R_0-\sqrt{x^2+y^2}\right)}{W}.
\end{equation}

\begin{figure}
	\centering
	\subfigure[]{
		\begin{minipage}{0.49\linewidth}
			\centering
			\includegraphics[width=3.0in]{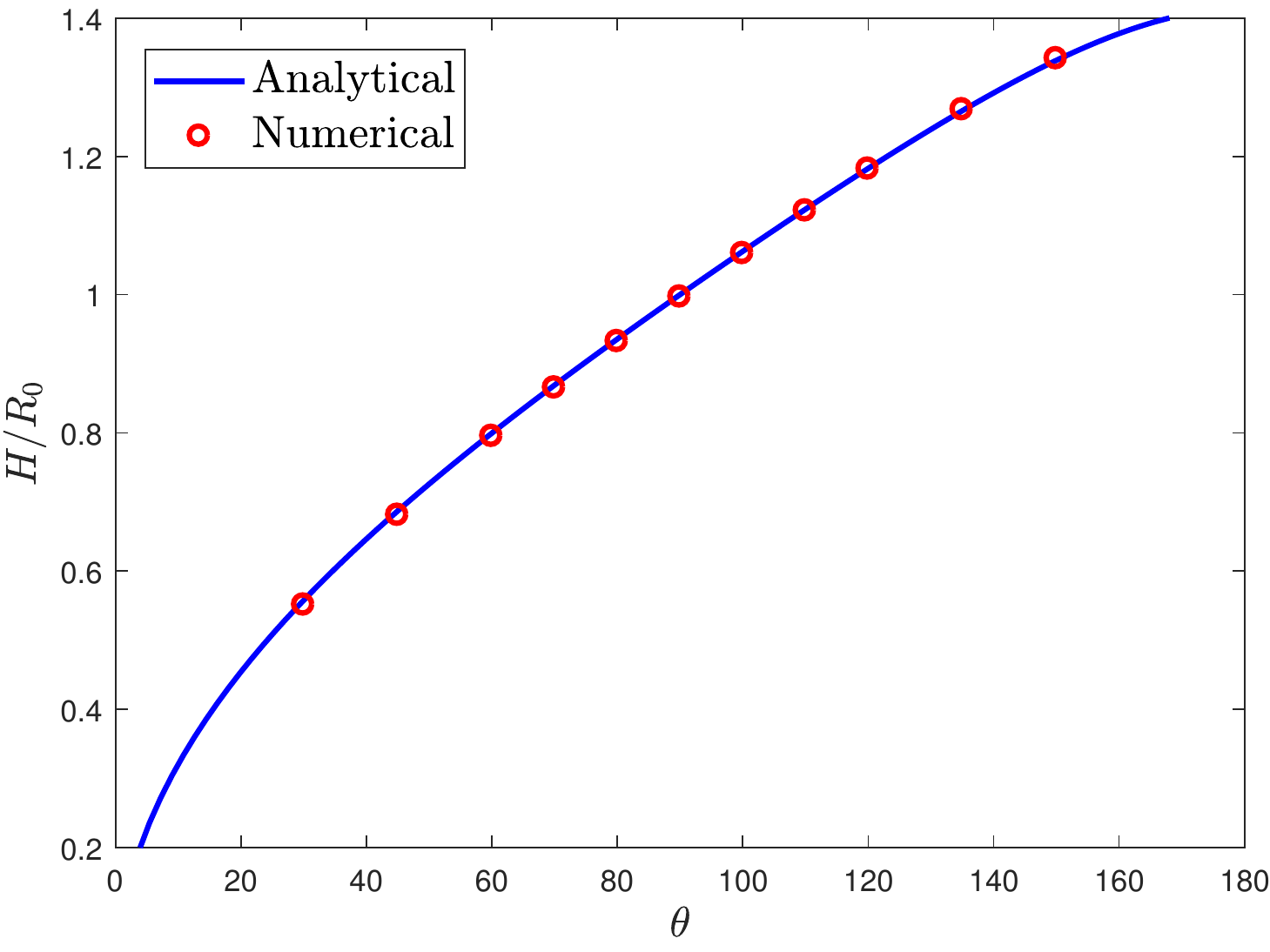}
			\label{fig-SpHd}
	\end{minipage}}	
	\subfigure[]{
		\begin{minipage}{0.49\linewidth}
			\centering
			\includegraphics[width=3.0in]{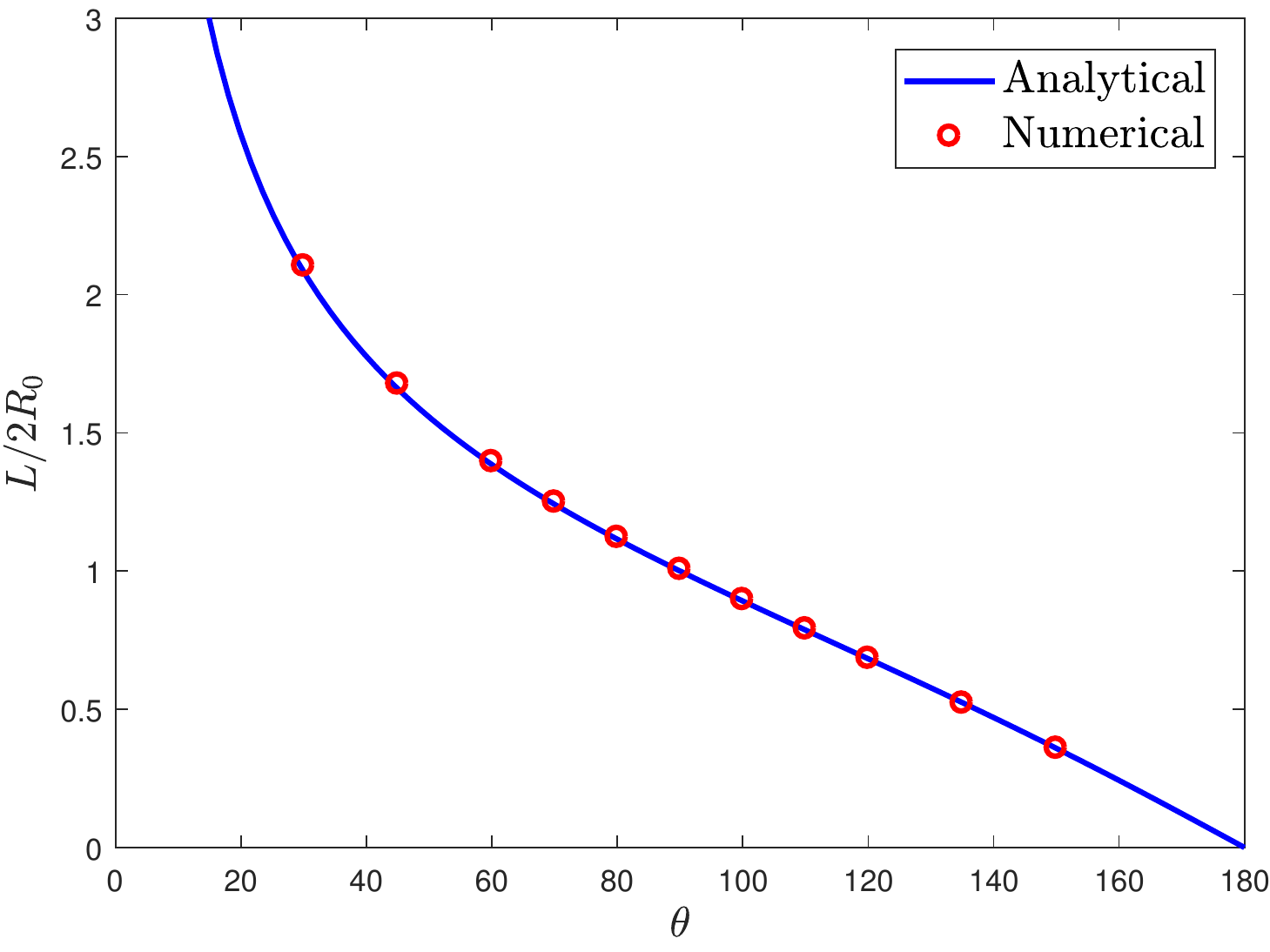}
			\label{fig-SpLd}
	\end{minipage}}	
	\caption{The geometric quantities of the spreading droplet at different contact angles [(a) The normalized height of the droplet, (b) The normalized spreading length of the droplet].}
	\label{fig-Sp-HLd}
\end{figure}	
\begin{table}
	\centering
	\caption{The contact angle of the droplet spreading on an ideal wall.}
	\begin{tabular}{ccccccccccccc}
		\toprule
		Theoretical     && 30\degree && 45\degree && 60\degree && 90\degree && 120\degree && 150\degree \\
		\midrule
		Numerical && 29.3\degree && 44.2\degree && 59.3\degree && 89.5\degree && 119.8\degree && 150.0\degree \\
		\bottomrule
	\end{tabular}
	\label{table-ContactAngle}
\end{table}
In addition, when the droplet contacts a solid, the wettability boundary condition must be considered. Here we adopt the following one \cite{Huang2015IJNMF,Liang2019PRE},
\begin{equation}\label{FEwetting}
	\mathbf{n}_w\cdot\nabla\phi=-\sqrt{\frac{2\beta}{k}}\left(\phi_A-\phi\right)\left(\phi-\phi_B\right)\cos\theta,
\end{equation}
where $\mathbf{n}_w$ is the unit vector normal to the solid surface, and has
the direction pointing from the solid toward the fluid. Like Ref. \cite{Liang2019PRE}, after a discretization of Eq.\,(\ref{FEwetting}), we have
\begin{equation}\label{Dwetting}
	a\phi_{x,w}^2+b\phi_{x,w}+c=0,
\end{equation}
where $a=\Delta x\sqrt{2\beta/k}\cos\theta=4\Delta x\cos\theta/\left[W\left(\phi_A-\phi_B\right)\right]$, $b=8/3-a\left(\phi_A+\phi_B\right)$, $c=a\phi_A\phi_B+\phi_{x,2}/3-3\phi_{x,1}$, and $\phi_{x,w}$ represents the unknown order parameter at the solid wall with the subscript $x$ denoting the horizontal coordinate.
Considering the case of $\phi_A=1$ and $\phi_B=0$, when $a>0$, Eq.\,(\ref{Dwetting}) only has one positive root $\left(-b+\sqrt{b^2-4ac}\right)/2a$ with $c<0$. On the other hand, if $a<0$, under the condition of  $\cos\theta\geq-1\geq-2W\left(\phi_A-\phi_B\right)/\left(3\Delta x\right)$ since $W/\Delta x\geq 1.5$ usually satisfied in phase-field method, the quadratic function $a\phi_{x,w}^2+b\phi_{x,w}+c$ has an axis of symmetry, 
\begin{equation}
	-\frac{b}{2a}=\frac{1}{2}-\frac{4}{3a}=\frac{1}{2}-\frac{W\left(\phi_A-\phi_B\right)}{3\Delta x\cos\theta}\geq1,
\end{equation} 
thus only the smaller root $\left(-b+\sqrt{b^2-4ac}\right)/2a$ of Eq.\,(\ref{Dwetting}) falls into the region $[0,1]$ with $c<0$. In general, the solution of Eq.\,(\ref{Dwetting}) that meets the accessing requirements should be $\left(-b+\sqrt{b^2-4ac}\right)/2a$ when $a\neq 0$.

In the following simulations, $\Delta x=\Delta t=1$, $W=5$, and some other physical parameters are set as $\rho_B=1$, $\nu_B=0.1$, $\sigma=0.2$ and $M_{\phi}=0.1$. Fig.\,\ref{fig-Sp} shows the equilibrium shapes of the droplet under different prescribed contact angles. From this figure, one can observe that the droplet can form different steady patterns on the solid wall, which critically depends on the specified value of the contact angle. From the equilibrium state of the droplet, one can measure the droplet spreading length $L$ on the solid surface and its height $H$ with the geometric relations. As seen from Fig.\,\ref{fig-Sp-HLd}, the numerical solutions are in good agreement with the following exact solutions,
\begin{equation}
	H=R-R\cos\theta,\quad L=2R\sin\theta,\quad R=R_0\sqrt{\frac{\pi/2}{\theta-\sin\theta\cos\theta}}.
\end{equation}

In addition, the numerical value of the contact angle can also be derived according to the geometrical relation $\theta=2\arctan\left(2H/L\right)$. The predicted contact angles with the present LB model are listed in Table \ref{table-ContactAngle}, from which one can find that the numerical results are close to the theoretical values.

\subsection{The Rayleigh-Taylor instability}\label{RTI}
The RTI is a fundamental interfacial instability that occurs when a heavier fluid is accelerated against a lighter one in the presence of a slight perturbation at the interface.
The physical problem we consider here is in a domain of $[-d/2,d/2]\times [-2d,2d]$ with the periodic boundary condition in horizontal direction and no-flux conditions on top and bottom boundaries. The problem consists of two layers of fluids with a heavy fluid (phase A) at the top and a light one (phase B) at the bottom, the initial interface is $h=-0.1d\cos(2\pi x/d)$. To ensure the physical variables to be smoothed across the interface, the initial profile of order parameter is set as
\begin{equation}
	\phi\left(x,y\right)=\frac{\phi_A+\phi_B}{2}+\frac{\phi_A-\phi_B}{2}\tanh\frac{2\left(y-h\right)}{W}.
\end{equation} 

To depict the RTI problem, the following dimensionless Reynolds number, P{\'e}clet number, and the Atwood number are used,
\begin{equation}
	Re=\frac{d\sqrt{gd}}{\nu_A},\quad Pe=\frac{\sqrt{gd}W}{M_{\phi}\beta\left(\phi_A-\phi_B\right)^2},\quad At=\frac{\rho_A-\rho_B}{\rho_A+\rho_B},
\end{equation}
where $g$ is the magnitude of the gravitational acceleration, the body force $\mathbf{G}=-\left[0,\rho-(\rho_A+\rho_B)/2\right]g$ is imposed on two fluids. 

\begin{figure}
	\centering
	\subfigure[]{
		\begin{minipage}{0.49\linewidth}
			\centering
			\includegraphics[width=3.0in]{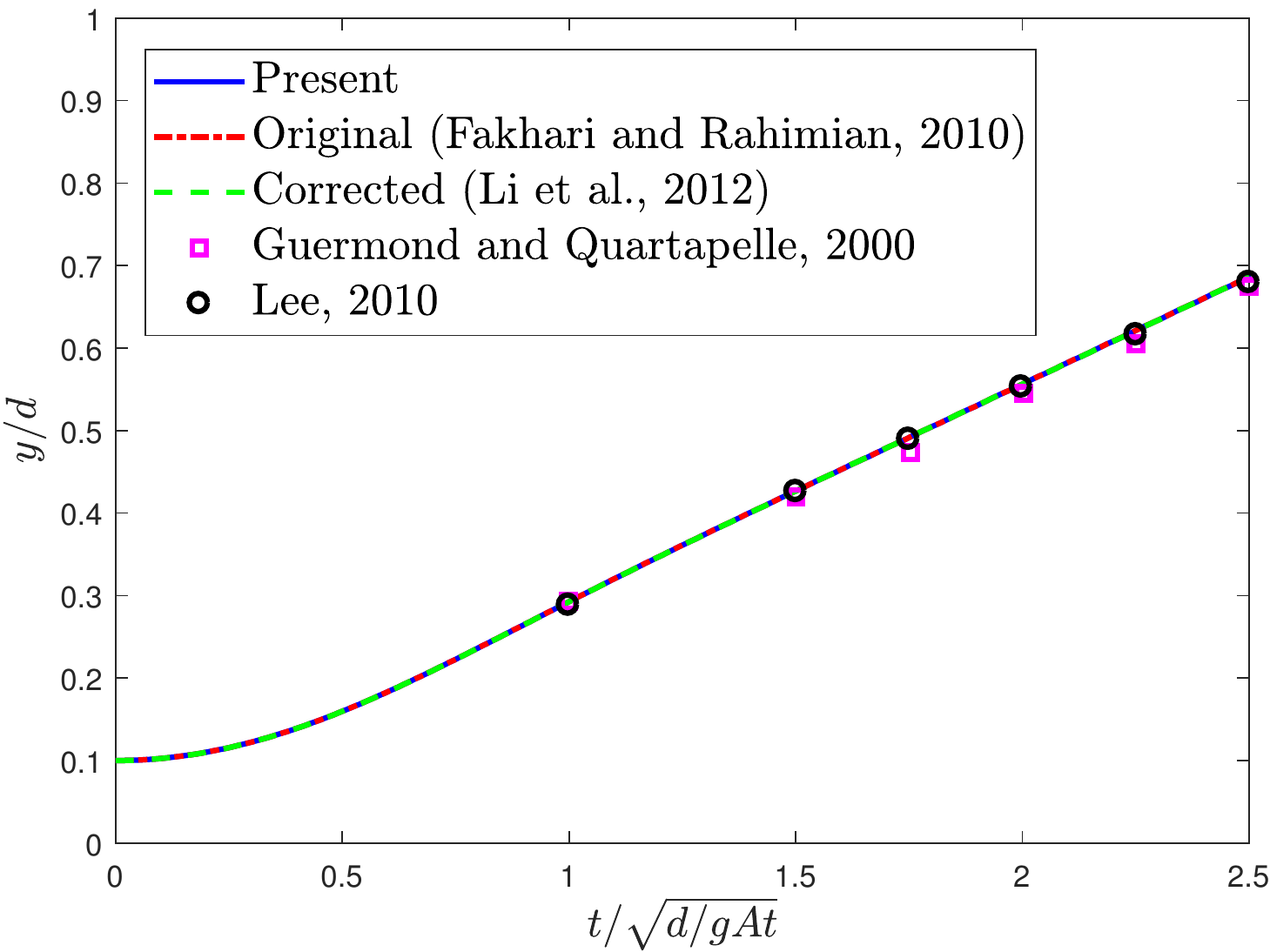}
			\label{fig-rising}
	\end{minipage}}	
	\subfigure[]{
		\begin{minipage}{0.49\linewidth}
			\centering
			\includegraphics[width=3.0in]{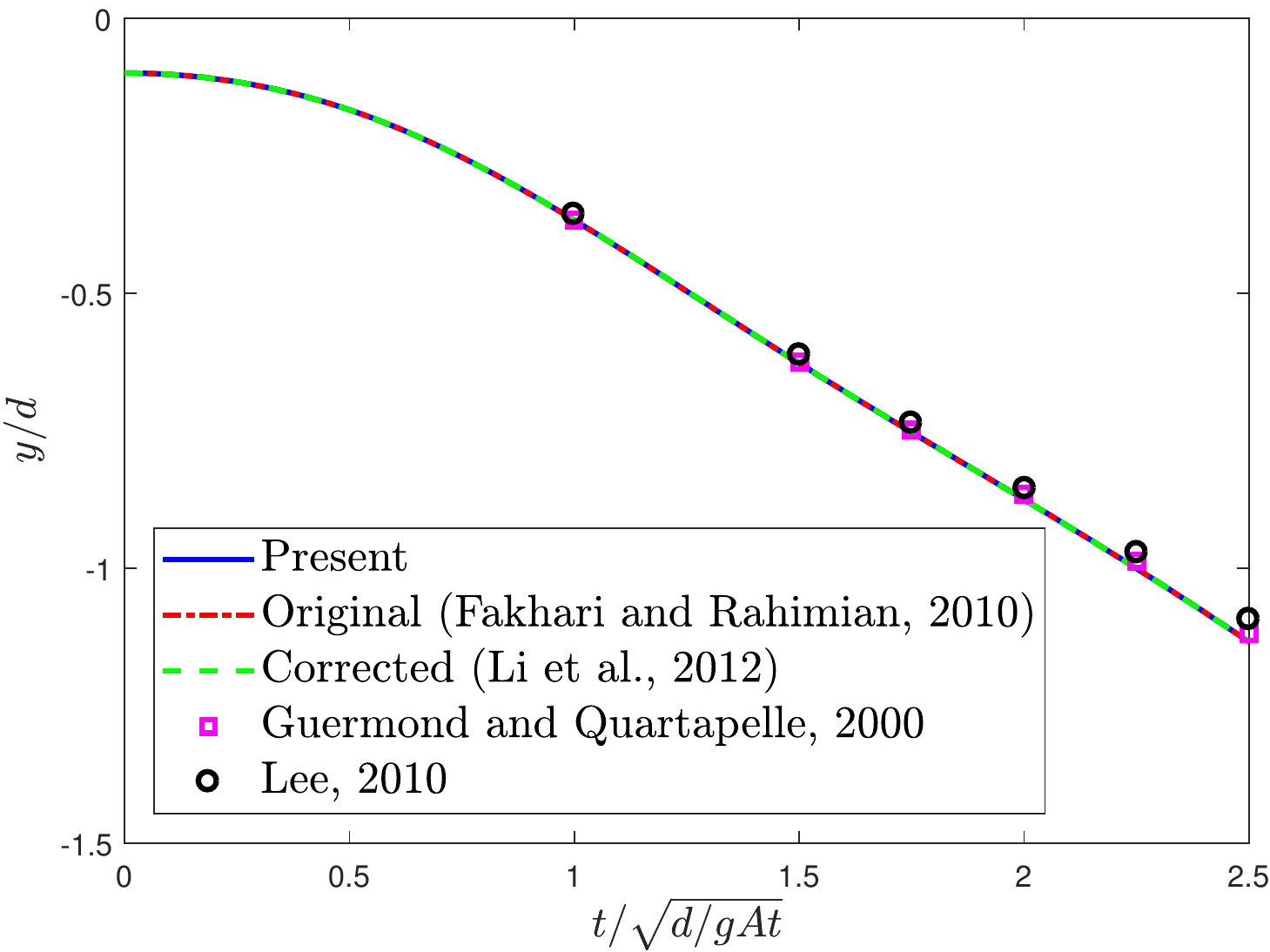}
			\label{fig-falling}
	\end{minipage}}	
	\caption{The normalized positions of the tips of the rising and falling fluids when $0\leq t/\sqrt{d/gAt}\leq 2.5$ [(a) The top of the rising fluid, (b) The bottom of the falling fluid].}
	\label{fig-RTI}
\end{figure}

\begin{figure}
	\centering
	\includegraphics[width=4.0in]{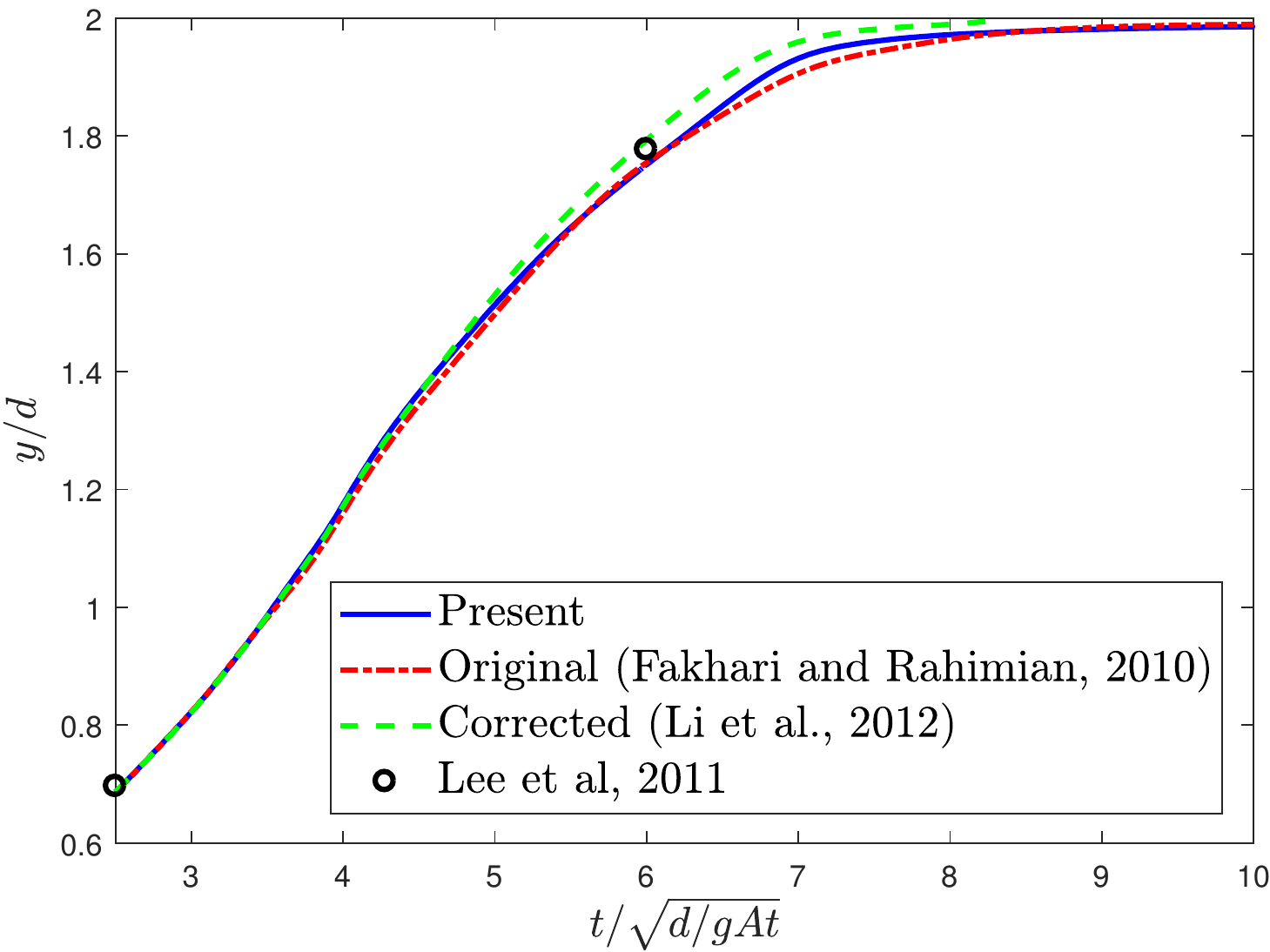}
	\caption{The normalized position of the top of the rising fluid when $2.5\leq t/\sqrt{d/gAt}\leq 10$.}
	\label{fig-rising6}
\end{figure}

In our simulations, the grid size is $150\times 600$, $\Delta t=1$, $s_f^3=s_f^5=0.9$, $s_g^4=s_g^6=1.7$, and some other physical parameters are fixed as $d=150$, $\sigma=5\times 10^{-5}$, $At=0.5$, $Pe=105$ and $Re=3000$ which are the same as those in the previous works \cite{Guermond2000JCP,Lee2010CF,Lee2011IJNME}. We presented the normalized positions of the top of the rising fluid and the bottom of the falling fluid in Fig.\,\ref{fig-RTI}. From this figure, one can find that there is an agreement among different LB models when $0\leq t/\sqrt{d/gAt}\leq 2.5$. However, as shown in Fig.\,\ref{fig-rising6}, some obvious differences are also observed when $2.5\leq t/\sqrt{d/gAt}\leq 10$, which may be caused by the different forms of the additional interfacial force (see the details in \ref{comparison}).

\subsection{Dam break}\label{Dambreak}
\begin{figure}
	\centering
	\includegraphics[width=4.0in]{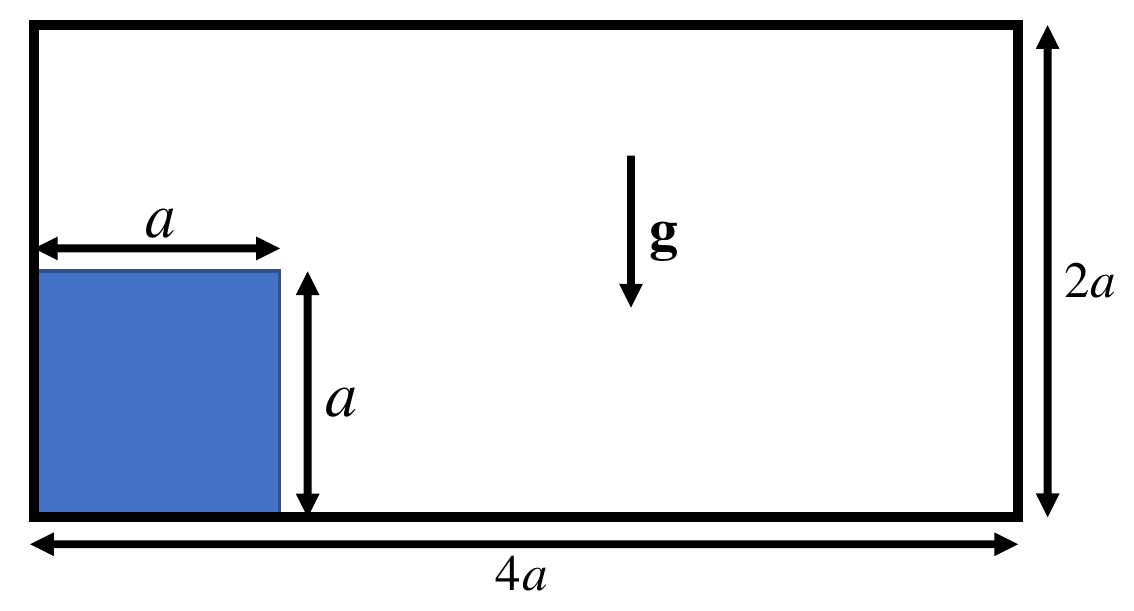}
	\caption{The configuration of the two-dimensional dam break.}
	\label{fig-DamDry}
\end{figure}

The last problem we considered is the dam break, which can be applied to test the ability of the present LB model in the study of the free surface flows with the high Reynolds numbers and large density ratios. With the time increasing, the water in the chamber will collapses, break and splash under the action of the gravity. The schematic of the problem is shown in Fig.\,\ref{fig-DamDry} where the computational domain is $Lx\times Ly=0.2\times 0.1$, the water is initially placed at the left bottom of the chamber with the length $a=0.05$.
For this problem, the order parameter is initialized by
\begin{equation}
	\phi\left(x,y\right)=\begin{cases}
		\frac{\phi_A+\phi_B}{2}+\frac{\phi_A-\phi_B}{2}\tanh\frac{2\left(a-y\right)}{W},\quad x\leq a-W, y\geq a-W,\\
		\frac{\phi_A+\phi_B}{2}+\frac{\phi_A-\phi_B}{2}\tanh\frac{2\left(a-x\right)}{W},\quad x\geq a-W, y\leq a-W,\\
		\frac{\phi_A+\phi_B}{2}+\frac{\phi_A-\phi_B}{2}\tanh\frac{2\left[W-\sqrt{\left(x-a+W\right)^2+\left(y-a+W\right)^2}\right]}{W},\quad x\geq a-W, y\geq a-W,\\
		\phi_A,\quad \text{otherwise}.\\
	\end{cases}
\end{equation}

In our simulations, $W=4\Delta x=5\times 10^{-4}$, $c=28$, $M_{\phi}=1\times 10^{-4}$, and some other physical parameters are $\rho_A=998.207$, $\rho_B=1.204$, $\sigma=7.28\times 10^{-3}$, $\mu_A=1.002\times 10^{-3}$, $\mu_B=1.78\times 10^{-5}$ and the gravity force $\mathbf{G}=\left(0,-\rho g\right)$, which are the material properties of water and air. Based on the values of these parameters, the Reynolds number defined by $Re=\rho_Aa\sqrt{ga}/\mu_A$ is 24655. 
For this problem, the inverse linear form of viscosity [Eq.\,(\ref{inverseNu})] is adopted. 

\begin{figure}
	\centering
	\subfigure[]{
		\begin{minipage}{0.3\linewidth}
			\centering
			\includegraphics[width=1.8in]{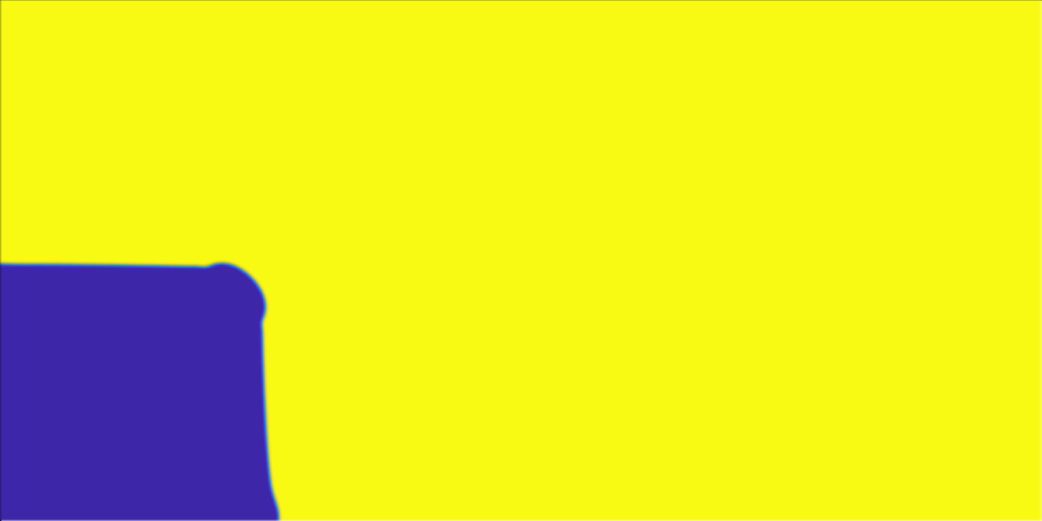}
			\label{fig-dam-t=0.25}
	\end{minipage}}	
	\subfigure[]{
		\begin{minipage}{0.3\linewidth}
			\centering
			\includegraphics[width=1.8in]{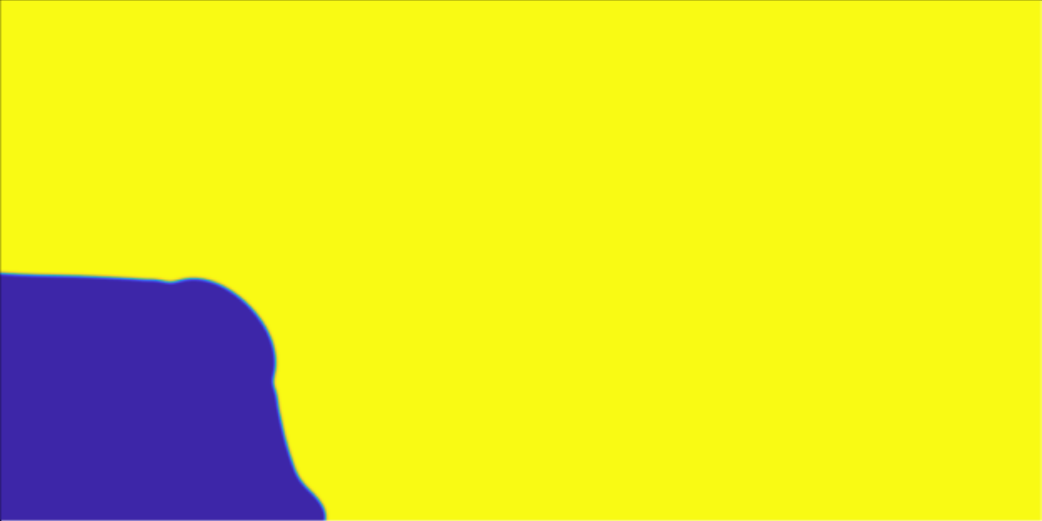}
			\label{fig-dam-t=0.5}
	\end{minipage}}		
	\subfigure[]{
		\begin{minipage}{0.3\linewidth}
			\centering
			\includegraphics[width=1.8in]{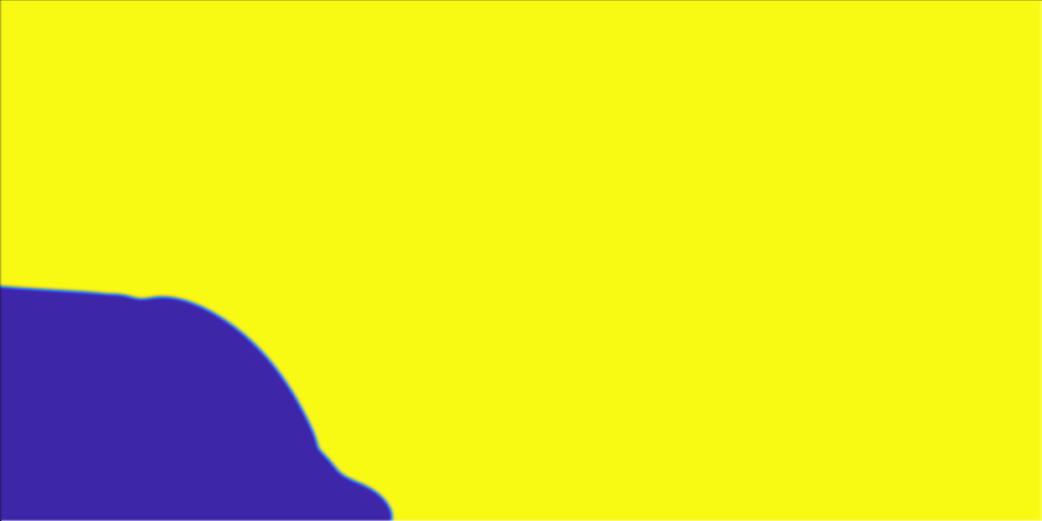}
			\label{fig-dam-t=0.75}
	\end{minipage}}	
	
	\subfigure[]{
		\begin{minipage}{0.3\linewidth}
			\centering
			\includegraphics[width=1.8in]{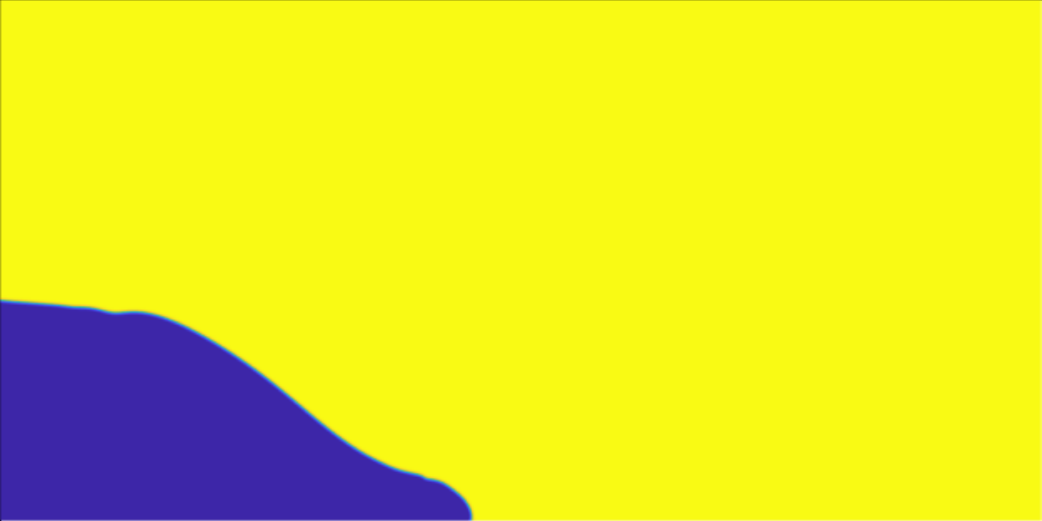}
			\label{fig-dam-t=1}
	\end{minipage}}	
	\subfigure[]{
		\begin{minipage}{0.3\linewidth}
			\centering
			\includegraphics[width=1.8in]{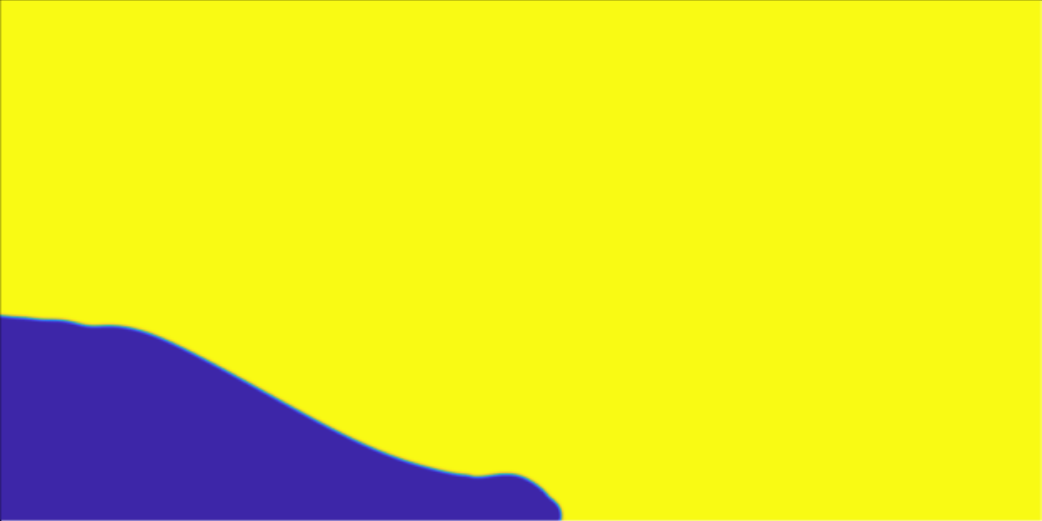}
			\label{fig-dam-t=1.25}
	\end{minipage}}	
	\subfigure[]{
		\begin{minipage}{0.3\linewidth}
			\centering
			\includegraphics[width=1.8in]{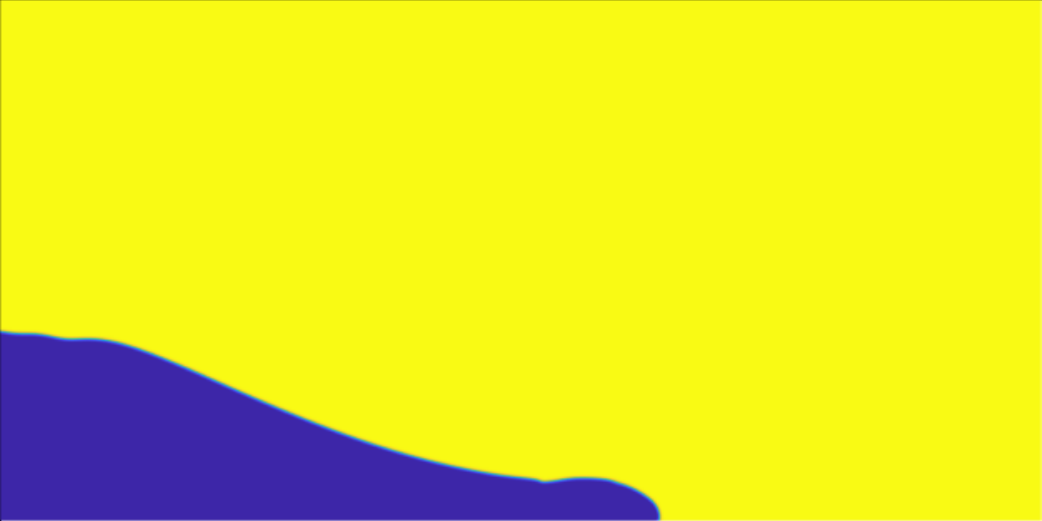}
			\label{fig-dam-t=1.5}
	\end{minipage}}	
	
	\subfigure[]{
		\begin{minipage}{0.3\linewidth}
			\centering
			\includegraphics[width=1.8in]{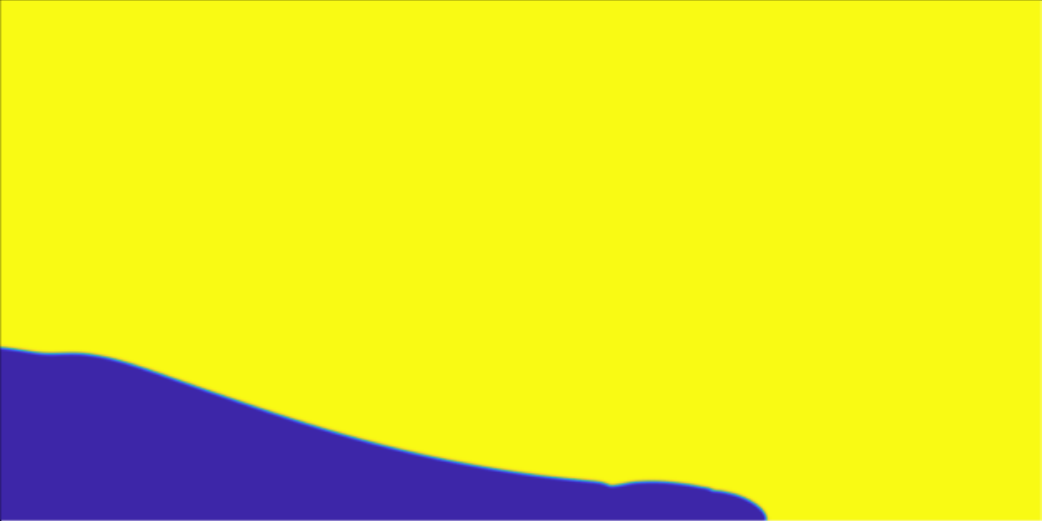}
			\label{fig-dam-t=1.75}
	\end{minipage}}	
	\subfigure[]{
		\begin{minipage}{0.3\linewidth}
			\centering
			\includegraphics[width=1.8in]{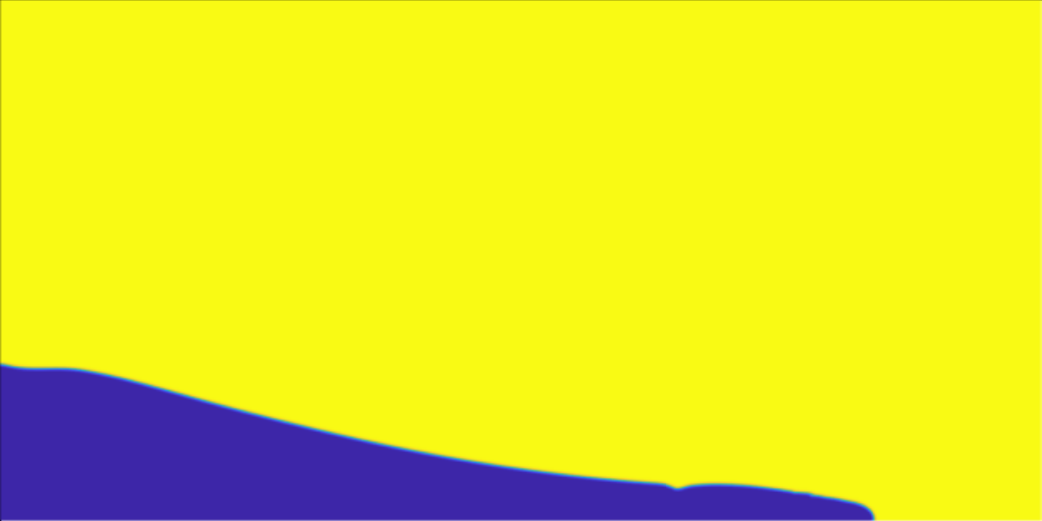}
			\label{fig-dam-t=2}
	\end{minipage}}	
	\subfigure[]{
		\begin{minipage}{0.3\linewidth}
			\centering
			\includegraphics[width=1.8in]{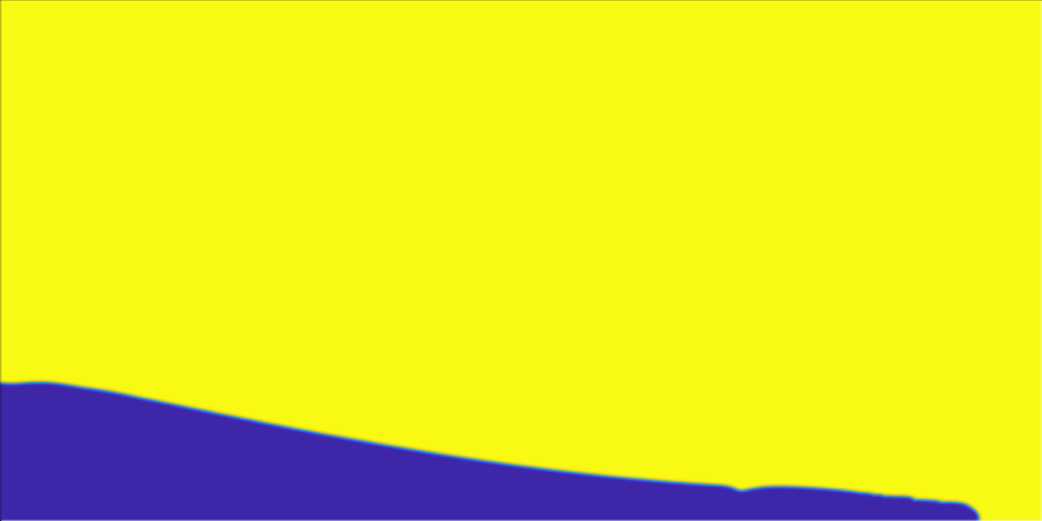}
			\label{fig-dam-t=2.25}
	\end{minipage}}	
	\caption{The snapshots of the dam break at different dimensionless time $T=t/\sqrt{a/g}$ [(a) $T=0.25$, (b) $T=0.5$, (c) $T=0.75$, (d) $T=1.0$, (e) $T=1.25$, (f) $T=1.5$, (g) $T=1.75$, (h) $T=2.0$, (i) $T=2.25$].}
	\label{fig-dam-T}
\end{figure}

\begin{figure}
	\centering
	\subfigure[]{
		\begin{minipage}{0.49\linewidth}
			\centering
			\includegraphics[width=3.0in]{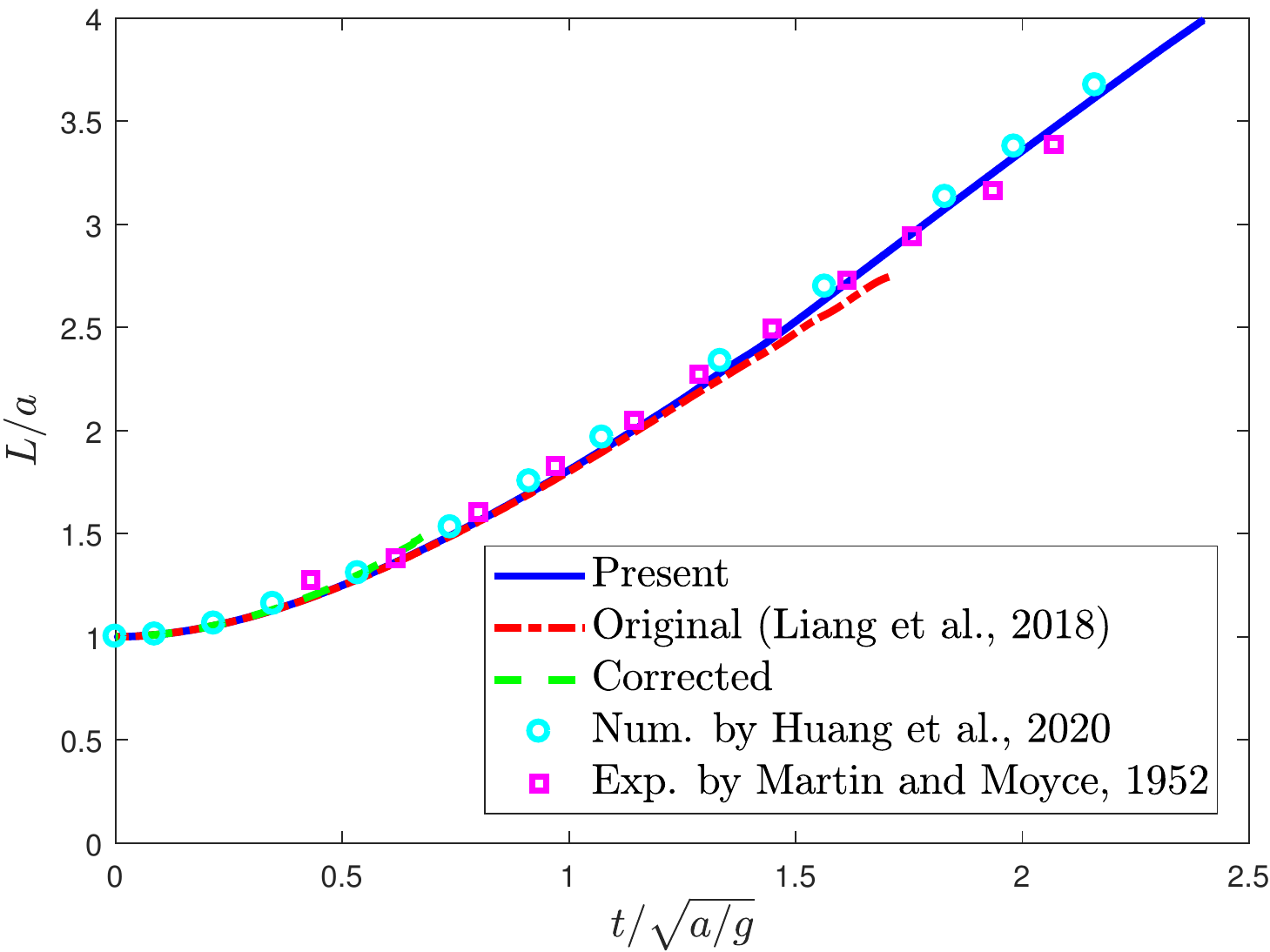}
			\label{fig-dam-L}
	\end{minipage}}	
	\subfigure[]{
		\begin{minipage}{0.49\linewidth}
			\centering
			\includegraphics[width=3.0in]{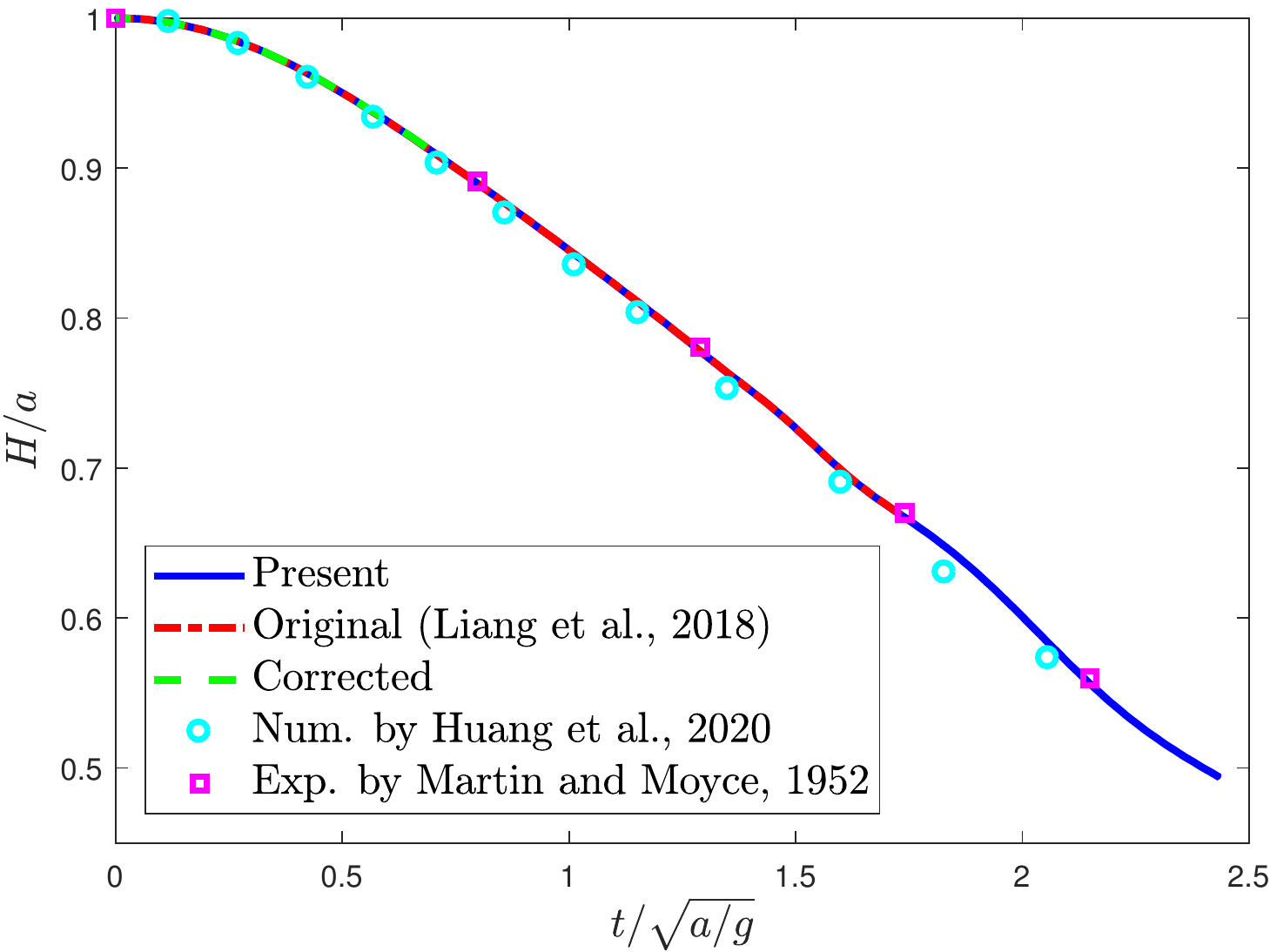}
			\label{fig-dam-H}
	\end{minipage}}	
	\caption{A comparison of the present LB model and some previous works. [(a) The normalized location of the front $L/a$, (b) The normalized location of the height $H/a$].}
	\label{fig-Dam-HL}
\end{figure}

The wettability is also considered, and the computational scheme is the same as that in Section \ref{DroSp}. The contact angle is first set as $90\degree$, and the snapshots of the dam break in time are shown in Fig.\,\ref{fig-dam-T}. As seen from this figure, the water collapses and spreads on the bottom wall under the action of the gravity until it impacts on the right wall. We also conducted a quantitative comparison of our numerical results with the previous experimental and numerical data in Fig.\,\ref{fig-Dam-HL}. As shown in this figure, the present LB model is more stable, and can capture the phenomena of dam break for a long time, compared to the original and corrected LB models. And also, the present results are in good agreement with those reported in Refs. \cite{Huang2020JCPa,Martin1952MPS}. 

\begin{figure}
	\centering
	\subfigure[]{
		\begin{minipage}{0.49\linewidth}
			\centering
			\includegraphics[width=3.0in]{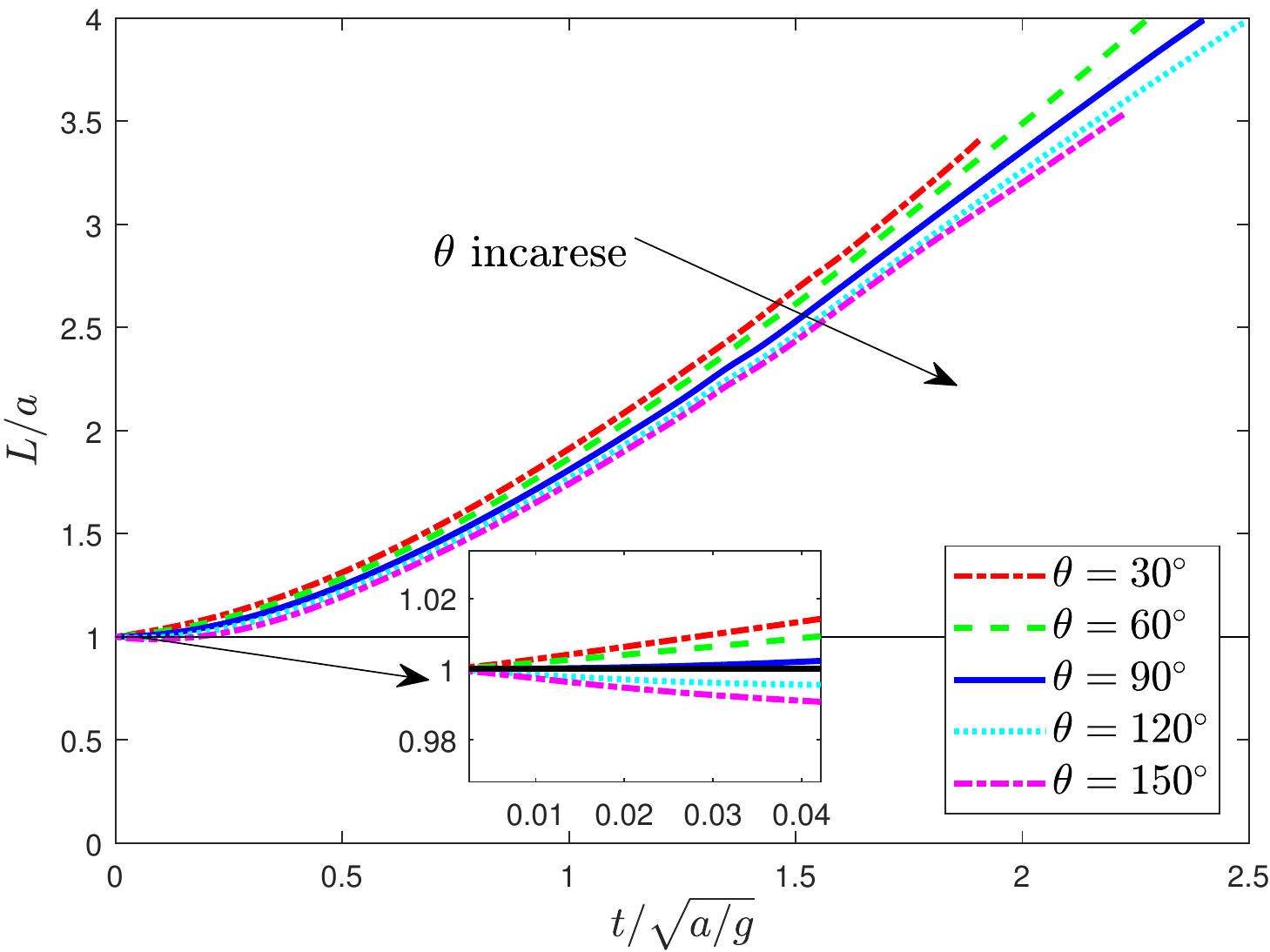}
			\label{fig-dam-thetaL}
	\end{minipage}}	
	\subfigure[]{
		\begin{minipage}{0.49\linewidth}
			\centering
			\includegraphics[width=3.0in]{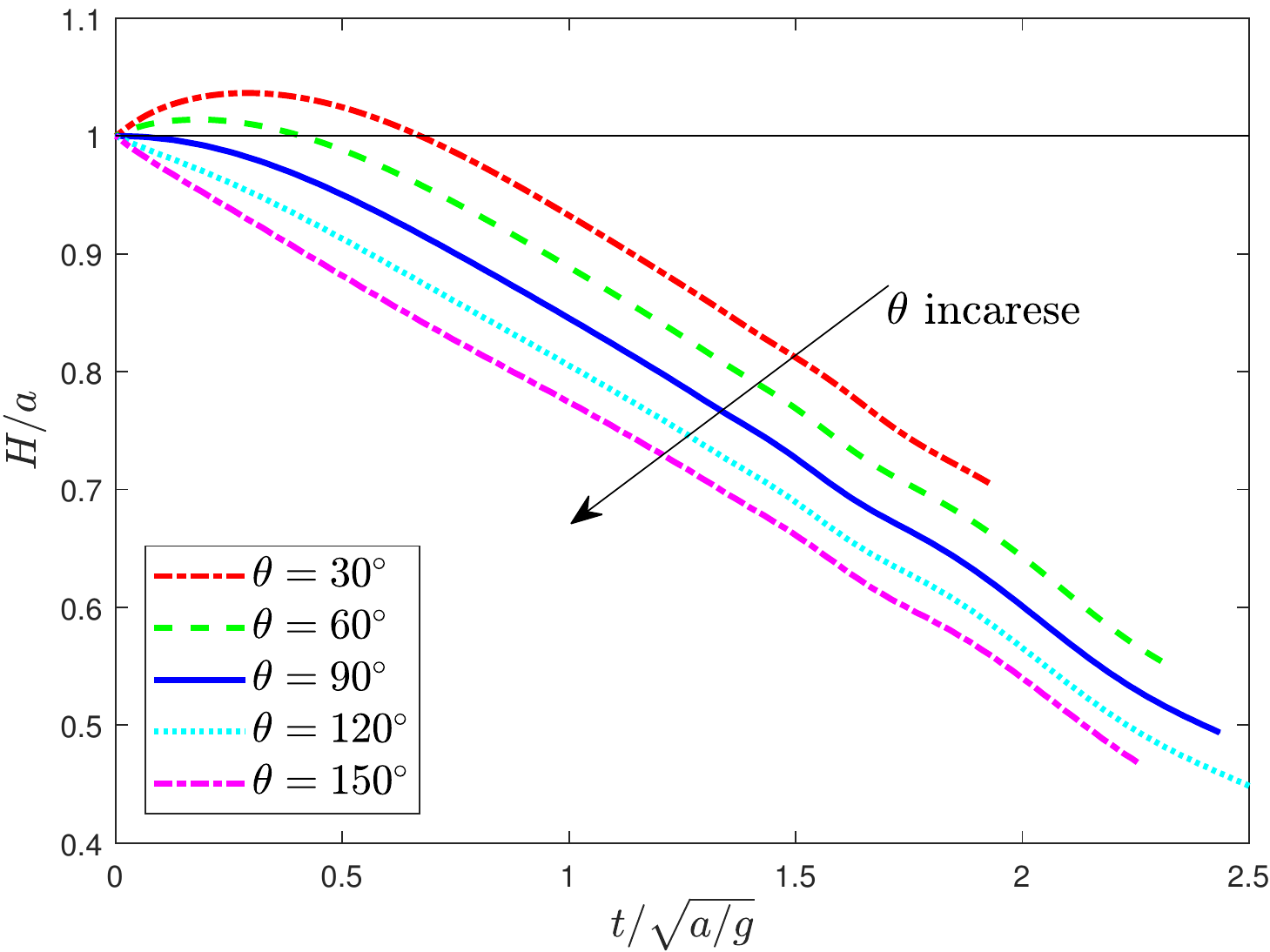}
			\label{fig-dam-thetaH}
	\end{minipage}}	
	\caption{A comparison of the results of the dam break at different contact angles [(a) The normalized location of the front $L/a$, (b) The normalized location of the height $H/a$].}
	\label{fig-Dam-theta}
\end{figure}

We further investigated the effect of the contact angle, and plotted the results in Fig.\,\ref{fig-Dam-theta}. When the contact angle increases from $30\degree$ to $150\degree$, the normalized locations of the front $L/a$ and the height $H/a$ both decrease due to the action of the gravity and wettability. Actually, the process of the dam break can be accelerated in the chamber with a large contact angle at the left wall while a small one at the bottom wall. In addition, at the initial stage, one can also observe that the location of the front changes in the opposite trend when contact angle is greater than $90\degree$, while the similar change of the location of the height occurs when contact angle is smaller than $90\degree$, which can be explained by the fact that the surface tension plays a dominated role at the initial stage instead of the gravity.    
\begin{figure}
	\centering
	\includegraphics[width=4.0in]{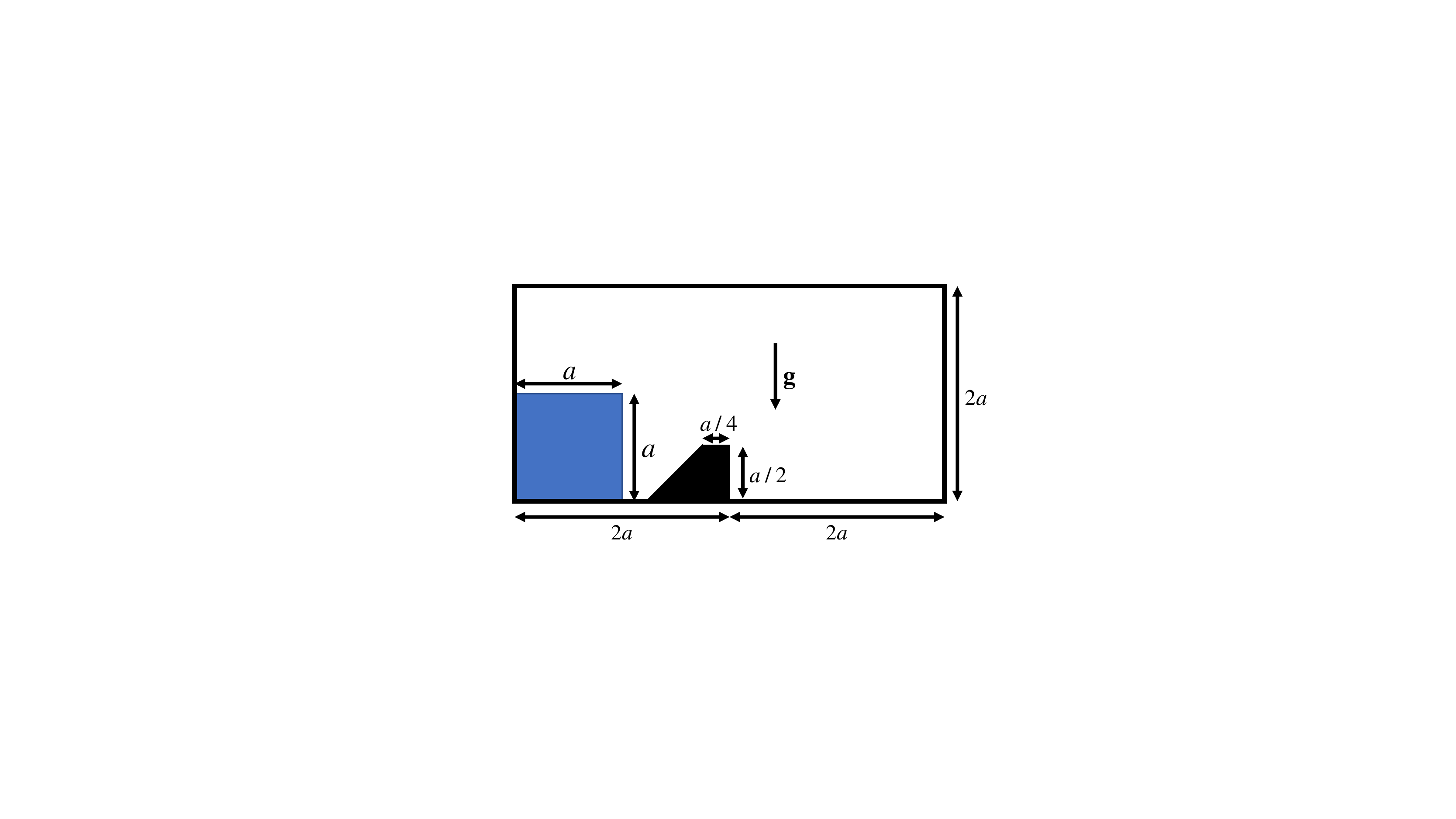}
	\caption{The configuration of the two-dimensional dam break with a trapezoidal obstacle.}
	\label{fig-DamTob}
\end{figure}
\begin{figure}
	\centering
	\subfigure[]{
		\begin{minipage}{0.3\linewidth}
			\centering
			\includegraphics[width=1.8in]{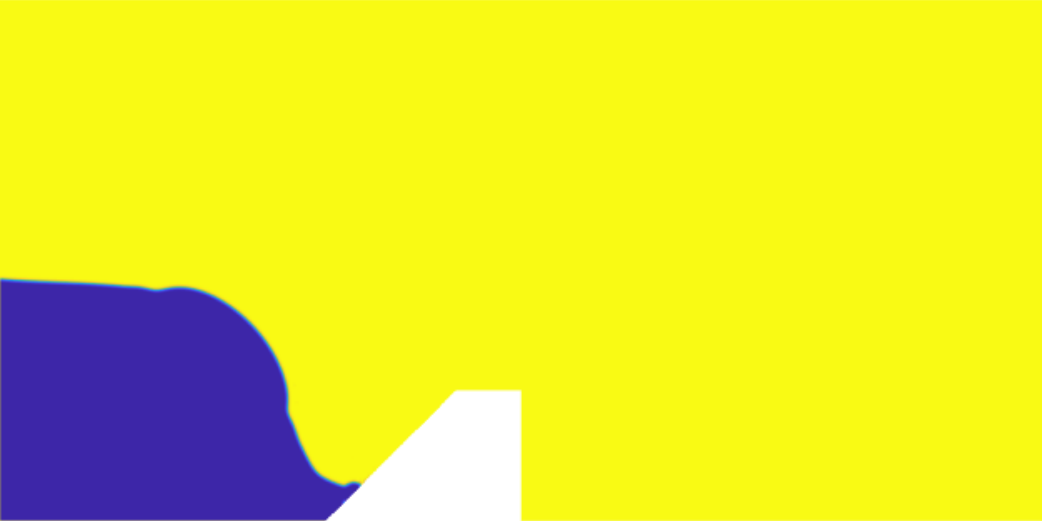}
			\label{fig-dam-original}
	\end{minipage}}	
	\subfigure[]{
		\begin{minipage}{0.3\linewidth}
			\centering
			\includegraphics[width=1.8in]{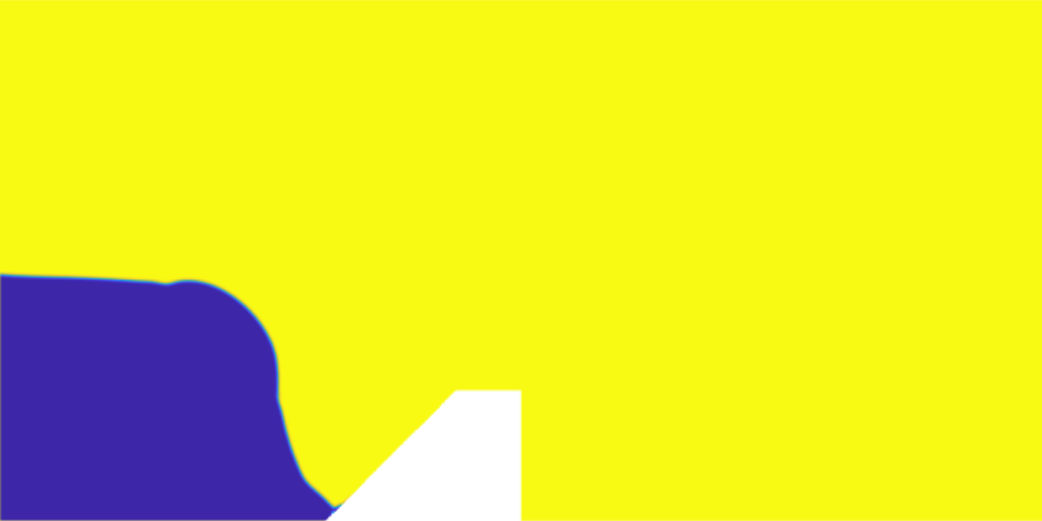}
			\label{fig-dam-corrected}
	\end{minipage}}		
	\subfigure[]{
		\begin{minipage}{0.3\linewidth}
			\centering
			\includegraphics[width=1.8in]{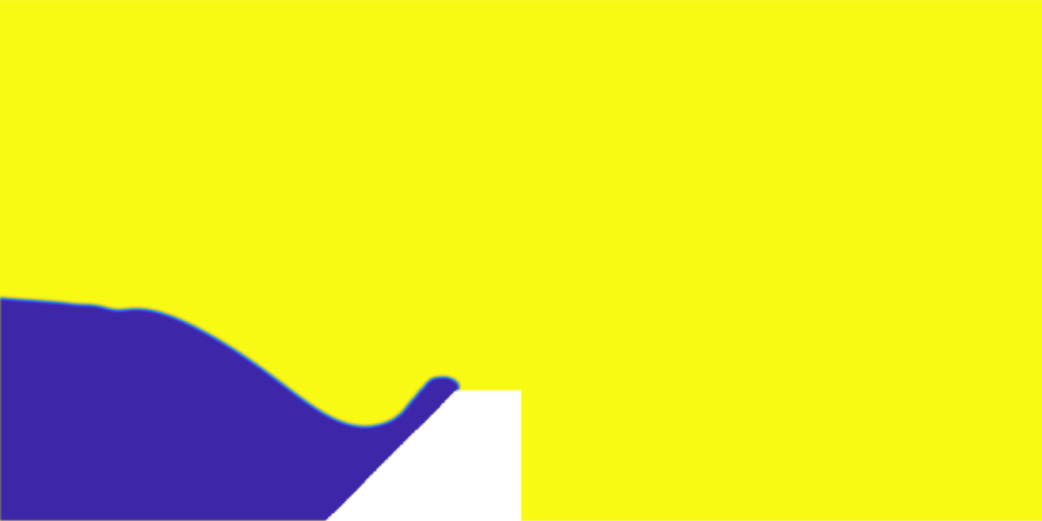}
			\label{fig-dam-Pret=0.75}
	\end{minipage}}	
	
	\subfigure[]{
		\begin{minipage}{0.3\linewidth}
			\centering
			\includegraphics[width=1.8in]{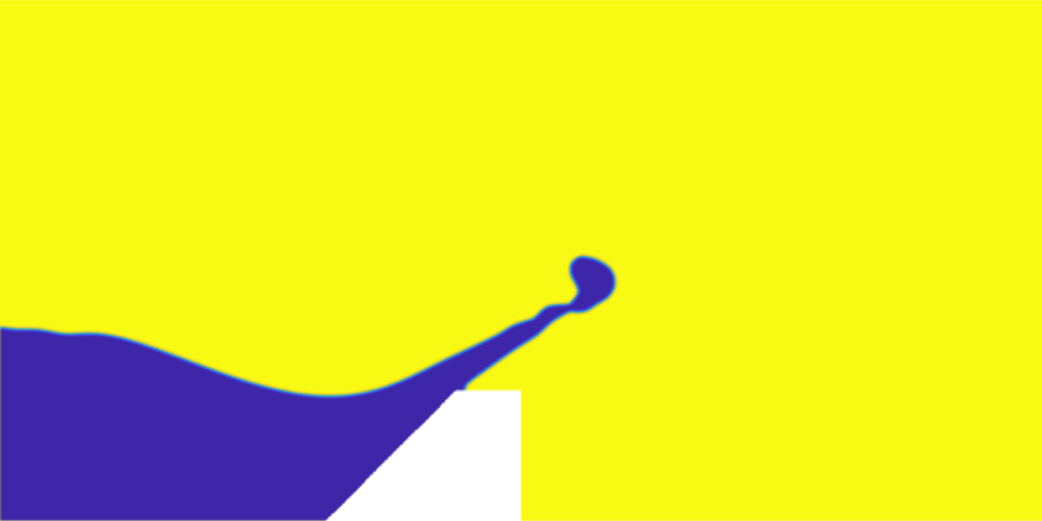}
			\label{fig-dam-Pret=1}
	\end{minipage}}	
	\subfigure[]{
		\begin{minipage}{0.3\linewidth}
			\centering
			\includegraphics[width=1.8in]{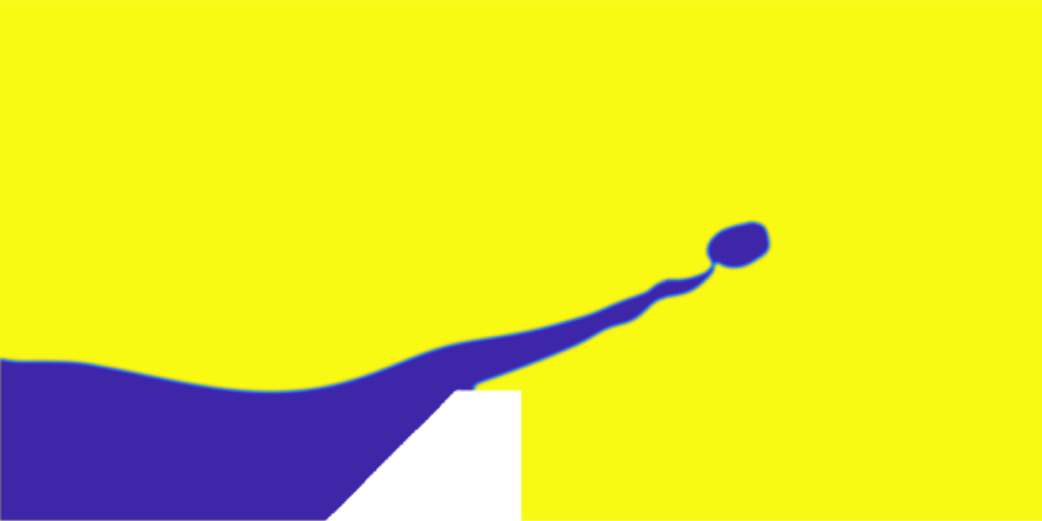}
			\label{fig-dam-Pret=1.25}
	\end{minipage}}	
	\subfigure[]{
		\begin{minipage}{0.3\linewidth}
			\centering
			\includegraphics[width=1.8in]{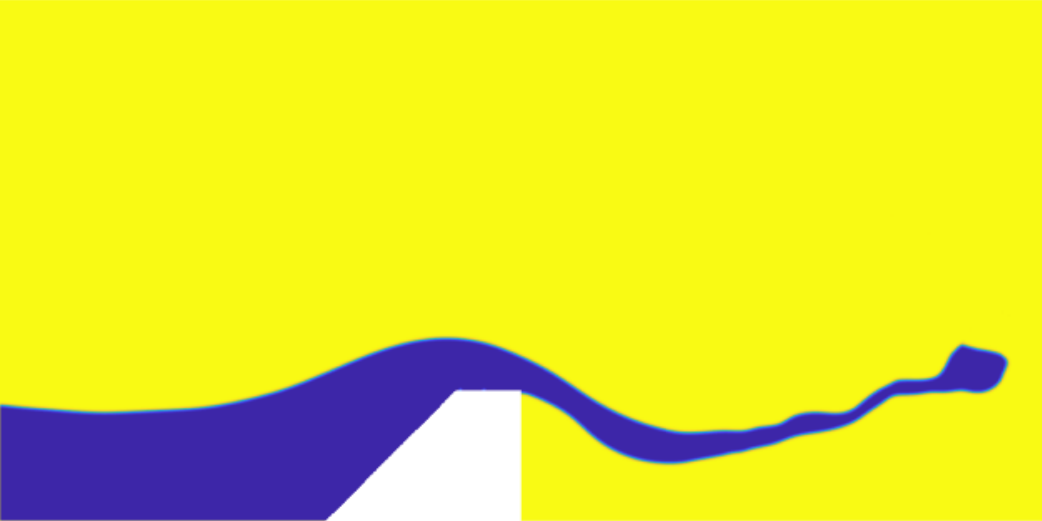}
			\label{fig-dam-Pret=1.5}
	\end{minipage}}	
	\caption{The results of original LB model (a) and corrected LB model (b) at the time before blow-up, present LB model at dimensionless time $T=t/\sqrt{a/g}$ [(c) $T=1.0$, (d) $T=1.5$, (e) $T=2.0$, (f) $T=3.0$].}
	\label{fig-dam-Tob}
\end{figure}

Finally, we studied the dam break with a trapezoidal solid obstacle on the bottom wall, and the configuration of the problem is shown in Fig.\,\ref{fig-DamTob} where the obstacle with the height length $a/2$ is placed at $Lx/2$ on the bottom of the chamber. We carried some simulations, and presented the snapshots of the dam break obtained by different LB models in Fig.\,\ref{fig-dam-Tob}. The results show that the original and corrected LB models [see Figs.\,\ref{fig-dam-original} and \ref{fig-dam-corrected}] are more unstable, and cannot simulate the phenomenon when the water rushes up along the slope, while the present LB model can [see Figs.\,\ref{fig-dam-Pret=0.75}-\ref{fig-dam-Pret=1.5}]. These results clearly illustrate that present LB model can capture the large deformation of the interface of two-phase flow problems. 

\section{Conclusion}\label{conclusion}
In this paper, a consistent and conservative phase-field based LB method with the MRT model is developed for the incompressible two-phase flows. Compared to the previous mathematical models considered in the phase-field LB method for the incompressible two-phase flows, the governing equations considered in this work satisfy three consistency conditions, and particularly, the mass flux in NS equations is reformulated to ensure the consistency of mass conservation. Actually, the continuity equation in present system is consistent to the one derived from the phase-filed equation such that it does not need to be solved, and moreover, a new additional interfacial force is obtained. The incompressibility and consistency of mass conservation are first evaluated by two statistical variables, and the results show that present LB model is better in terms of the consistency of mass conservation, compared to the original and corrected LB models. Then two benchmark problems, i.e., the layered Poiseuille flow and a droplet spreading on an ideal wall, are then used to test the accuracy of the present LB model. Finally, the RTI and the dam break problems with the high Reynolds numbers and large density ratios are considered, and the results illustrate that the present LB model is more stable for such complex problems.    

\section*{Acknowledgments}
This work was supported by the National Natural Science Foundation of China (Grants No. 12072127 and No. 51836003), and the Fundamental Research Funds for the Central Universities, HUST (No. 2021JYCXJJ010).

\appendix
\section{A comparison of two additional interfacial forces}\label{comparison}
In this appendix, we will present a comparison of the additional interfacial force between the present consistent and conservative LB method and the one in Ref. \cite{Li2012PRE}.
In the present LB model, the mass flux is $\mathbf{m}=\rho\mathbf{u}+\mathbf{m}^{\phi C}$, and the momentum equation is given by
\begin{equation}
	\frac{\partial\left(\rho\mathbf{u}\right)}{\partial t}+\nabla\cdot\left(\rho\mathbf{uu}+\mathbf{m}^{\phi C}\mathbf{u}\right)=-\nabla p+\nabla\cdot\mu\left[\nabla\mathbf{u}+\left(\nabla\mathbf{u}\right)^T\right]+\mu_{\phi}\nabla\phi+\mathbf{S}_{\mathbf{u}},
\end{equation}
which can also be written into another form,
\begin{equation}\label{NS-Fa1}
	\frac{\partial\left(\rho\mathbf{u}\right)}{\partial t}+\nabla\cdot\left(\rho\mathbf{uu}\right)=-\nabla p+\nabla\cdot\mu\left[\nabla\mathbf{u}+\left(\nabla\mathbf{u}\right)^T\right]+\mu_{\phi}\nabla\phi+\mathbf{F}_{p}+\mathbf{S}_{\mathbf{u}},
\end{equation}
where $\mathbf{F}_{p}=-\nabla\cdot\left(\mathbf{m}^{\phi C}\mathbf{u}\right)=-\mathbf{u}\nabla\cdot\mathbf{m}^{\phi C}-\mathbf{m}^{\phi C}\cdot\nabla\mathbf{u}$. Actually, one can also develop the LB model for Eq.\,(\ref{NS-Fa1}), same as the one in Ref. \cite{Wang2019Capillarity} except the form of the total external force, however, the computational scheme of velocity is implicit, and a special treatment is needed.

In the previous LB model \cite{Li2012PRE}, if we take $\mathbf{m}=\rho\mathbf{u}$ and with the help of Eq.\,(\ref{PFrho}), one can obtain
\begin{equation}
	\begin{aligned}
		\frac{\partial\left(\rho\mathbf{u}\right)}{\partial t}+\nabla\cdot\left(\rho\mathbf{uu}\right)
		=&\rho\left(\frac{\partial\mathbf{u}}{\partial t}+\mathbf{u}\cdot\nabla\mathbf{u}\right)+\mathbf{u}\left[\frac{\partial\rho}{\partial t}+\nabla\cdot\left(\rho\mathbf{u}\right)\right]\\
		=&\rho\left(\frac{\partial\mathbf{u}}{\partial t}+\mathbf{u}\cdot\nabla\mathbf{u}\right)-\mathbf{u}\nabla\cdot\mathbf{m}^{\phi C}.
	\end{aligned}	
\end{equation}
According to above equation, the original momentum equation
\begin{equation}
	\rho\left(\frac{\partial\mathbf{u}}{\partial t}+\mathbf{u}\cdot\nabla\mathbf{u}\right)=-\nabla p+\nabla\cdot\mu\left[\nabla\mathbf{u}+\left(\nabla\mathbf{u}\right)^T\right]+\mu_{\phi}\nabla\phi+\mathbf{S}_{\mathbf{u}},
\end{equation} 
can be rewritten as a conservative form,
\begin{equation}\label{NS2-add}
	\frac{\partial\left(\rho\mathbf{u}\right)}{\partial t}+\nabla\cdot\left(\rho\mathbf{uu}\right)=-\nabla p+\nabla\cdot\mu\left[\nabla\mathbf{u}+\left(\nabla\mathbf{u}\right)^T\right]+\mu_{\phi}\nabla\phi+\mathbf{F}_{a}+\mathbf{S}_{\mathbf{u}},
\end{equation}
where $\mathbf{F}_{a}=-\mathbf{u}\nabla\cdot\mathbf{m}^{\phi C}=\frac{d\rho}{d\phi}\mathbf{u}\nabla\cdot M_{\phi}\mathbf{D}\left(\phi\right)$, which is the term introduced to eliminate an artificial interfacial force \cite{Li2012PRE}. For CH equation, $\mathbf{F}_{a}=\frac{d\rho}{d\phi}\mathbf{u}\nabla\cdot\left(M_{\phi}\nabla\mu_{\phi}\right)$, while for AC equation, $\mathbf{F}_{a}=\frac{d\rho}{d\phi}\mathbf{u}\nabla\cdot M_{\phi}\left(\nabla\phi-\lambda\mathbf{n}\right)$.
It should be noted that the conservative forms of Eqs.\,(\ref{PhaseField}), (\ref{NS1}) and (\ref{NS2-add}) cannot satisfy the consistency of mass conservation. 

\section{The simplification of term $\partial_t\left(\rho\mathbf{u}\mathbf{u}\right)$}\label{Prhouu}
With the help of  $\partial_tp=O(Ma^2)$, the term $\partial_t\left(\rho\mathbf{u}\mathbf{u}\right)$ can be simplified by
\begin{equation}\label{ptrhouu}
	\begin{aligned}
		\partial_t\left(\rho u_{\alpha}u_{\beta}\right)=&u_{\alpha}\partial_t\left(\rho u_{\beta}\right)+ u_{\beta}\partial_t\left(\rho u_{\alpha}\right)-u_{\alpha}u_{\beta}\partial_t\rho\\
		=&u_{\alpha}\left[F_{\beta}-\partial_{\gamma}\left(\rho u_{\beta}u_{\gamma}+m_{\beta}^{\phi C}u_{\gamma}+p\delta_{\beta\gamma}\right)\right]+u_{\beta}\left[F_{\alpha}-\partial_{\gamma}\left(\rho u_{\alpha}u_{\gamma}+m_{\alpha}^{\phi C}u_{\gamma}+p\delta_{\alpha\gamma}\right)\right]\\
		&-u_{\alpha}u_{\beta}\partial_t\rho+O(Ma\Delta t)\\
		=&u_{\alpha}F_{\beta}-u_{\alpha}\partial_{\gamma}\left(\rho u_{\beta}u_{\gamma}\right)-u_{\alpha}\partial_{\gamma}\left(m_{\beta}^{\phi C}u_{\gamma}\right)-u_{\alpha}\partial_{1\beta}p\\
		&+u_{\beta}F_{\alpha}-u_{\beta}\partial_{\gamma}\left(\rho u_{\alpha}u_{\gamma}\right)-u_{\beta}\partial_{\gamma}\left(m_{\alpha}^{\phi C}u_{\gamma}\right)-u_{\beta}\partial_{\alpha}p-u_{\alpha}u_{\beta}\partial_t\rho+O(Ma\Delta t)\\
		=&u_{\alpha}F_{\beta}+u_{\beta}F_{\alpha}-2u_{\alpha}u_{\beta}\partial_{\gamma}\left(\rho u_{\gamma}\right)-\rho u_{\alpha}u_{\gamma}\partial_{\gamma}u_{\beta}-\rho u_{\beta}u_{\gamma}\partial_{\gamma}u_{\alpha}-u_{\alpha}\partial_{\beta}p-u_{\beta}\partial_{\alpha}p\\
		&-u_{\alpha}u_{\gamma}\partial_{\gamma}m_{\beta}^{\phi C}-u_{\beta}u_{\gamma}\partial_{\gamma}m_{\alpha}^{\phi C}-u_{\alpha}u_{\beta}\partial_t\rho+O(Ma\Delta t)\\
		=&u_{\alpha}F_{\beta}+u_{\beta}F_{\alpha}+O(Ma\Delta t+Ma^2).
	\end{aligned}
\end{equation}

\section{The moments of the D2Q9 lattice model}\label{D2Q9}
In the D2Q9 lattice model, $c_s^2=c^2/3$ with $c=\Delta x/\Delta t$. If the following orthogonal transformation matrix is considered,
\begin{equation}\label{M1}
	\mathbf{M}=\begin{pmatrix}
		1 &  1 &  1 &  1 &  1 &  1 &  1 &  1 &  1\\
		-4 & -1 & -1 & -1 & -1 &  2 &  2 &  2 &  2\\
		4 & -2 & -2 & -2 & -2 &  1 &  1 &  1 &  1\\
		0 &  1 &  0 & -1 &  0 &  1 & -1 & -1 &  1\\
		0 & -2 &  0 &  2 &  0 &  1 & -1 & -1 &  1\\
		0 &  0 &  1 &  0 & -1 &  1 &  1 & -1 & -1\\
		0 &  0 & -2 &  0 &  2 &  1 &  1 & -1 & -1\\
		0 &  1 & -1 &  1 & -1 &  0 &  0 &  0 &  0\\
		0 &  0 &  0 &  0 &  0 &  1 & -1 &  1 & -1\\
	\end{pmatrix},
\end{equation}
one can get $s_g^3=s_g^5=s_1$, $s_g^7=s_g^8=s_2$, the parameters in the computation of pressure are $H=\frac{s_g^2-s_g^0}{9s_g^0s_g^2}$, $J=\frac{2s_g^1+s_g^2+s_g^1s_g^2}{3c^2s_g^1s_g^2}$, $K=\frac{1}{3c^2}$, and the moments of distribution functions are given by
\begin{equation}\label{MgMF1}
	\mathbf{m}_g=\begin{pmatrix}
		\rho_0\\
		-4\rho_0+\frac{6p+3\rho\mathbf{u}\cdot\mathbf{u}+3\mathbf{u}\cdot\mathbf{m}^{\phi C}}{c^2}\\
		4\rho_0-\frac{9p+3\rho\mathbf{u}\cdot\mathbf{u}+3\mathbf{u}\cdot\mathbf{m}^{\phi C}}{c^2}\\
		\frac{\rho u}{c}\\
		-\frac{\rho u}{c}\\
		\frac{\rho v}{c}\\
		-\frac{\rho v}{c}\\
		\frac{um_x^{\phi C}-vm_y^{\phi C}+\rho\left(u^2-v^2\right)}{c^2}\\
		\frac{um_y^{\phi C}+vm_x^{\phi C}+2\rho uv}{2c^2}\\
	\end{pmatrix},\quad
	\mathbf{m}_F=\begin{pmatrix}
		\mathbf{u}\cdot\nabla\rho\\
		\frac{3\partial_{t}\left(\mathbf{u}\cdot\mathbf{m}^{\phi C}\right)+6\mathbf{u}\cdot\mathbf{F}}{c^2}\\
		-\mathbf{u}\cdot\nabla\rho-\frac{3\partial_{t}\left(\mathbf{u}\cdot\mathbf{m}^{\phi C}\right)+6\mathbf{u}\cdot\mathbf{F}}{c^2}\\
		\frac{F_x}{c}\\
		-\frac{F_x}{c}\\
		\frac{F_y}{c}\\
		-\frac{F_y}{c}\\
		\frac{2}{3}\left(u\partial_x\rho-v\partial_y\rho\right)+\frac{2\left(uF_x-vF_y\right)+\partial_t\left(um_x^{\phi C}-vm_y^{\phi C}\right)}{c^2}\\
		\frac{1}{3}\left(u\partial_y\rho+v\partial_x\rho\right)+\frac{2\left(uF_y+vF_x\right)+\partial_t\left(um_y^{\phi C}+vm_x^{\phi C}\right)}{2c^2}\\
	\end{pmatrix},
\end{equation}
where $\mathbf{u}=\left(u,v\right)$, $\mathbf{m}^{\phi C}=\left(m_x^{\phi C},m_y^{\phi C}\right)$, $\mathbf{F}=\left(F_x,F_y\right)$ and $\nabla\rho=\left(\partial_x\rho,\partial_y\rho\right)$.

For the CH equation, $\mathbf{m}^{\phi C}=-\frac{d\rho}{d\phi}M_{\phi}\nabla\mu_{\phi}$, the moments are given by
\begin{equation}\label{MfMRCH1}
	\begin{aligned}
		&\mathbf{m}_f=\left(\phi,-4\phi+2\eta\mu_{\phi},4\phi-3\eta\mu_{\phi},\frac{\phi u}{c},-\frac{\phi u}{c},\frac{\phi v}{c},-\frac{\phi v}{c},0,0\right)^T,\\
		&\mathbf{m}_R=\left(0,0,0,\frac{\partial_t\phi u}{c},-\frac{\partial_t\phi u}{c},\frac{\partial_t\phi v}{c},-\frac{\partial_t\phi v}{c},0,0\right)^T,
	\end{aligned}
\end{equation}
where $\eta$ is an adjustable parameter, and is set as 1 by default. 
For AC equation, we have $\mathbf{m}^{\phi C}=-\frac{d\rho}{d\phi}M_{\phi}\left(\nabla\phi-\lambda\mathbf{n}\right)$ with $\mathbf{n}=\left(n_x,n_y\right)$, 
the moments of distribution functions can be determined by
\begin{equation}\label{MfMRAC1}
	\begin{aligned}
		&\mathbf{m}_f=\left(\phi,-4\phi+2\eta\phi,4\phi-3\eta\phi,\frac{\phi u}{c},-\frac{\phi u}{c},\frac{\phi v}{c},-\frac{\phi v}{c},0,0\right)^T,\\
		&\mathbf{m}_R=\left(0,0,0,\frac{c\lambda n_x}{3}+\frac{\partial_t\phi u}{c},-\frac{c\lambda n_x}{3}-\frac{\partial_t\phi u}{c},\frac{c\lambda n_y}{3}+\frac{\partial_t\phi v}{c},-\frac{c\lambda n_y}{3}-\frac{\partial_t\phi v}{c},0,0\right)^T.
	\end{aligned}
\end{equation}

If the transformation matrix $\mathbf{M}$ has the following form,
\begin{equation}
	\mathbf{M}=\begin{pmatrix}
		1 &  1 &  1 &  1 &  1 &  1 &  1 &  1 &  1\\
		0 &  1 &  0 & -1 &  0 &  1 & -1 & -1 &  1\\
		0 &  0 &  1 &  0 & -1 &  1 &  1 & -1 & -1\\
		0 &  1 &  0 &  1 &  0 &  1 &  1 &  1 &  1\\
		0 &  0 &  0 &  0 &  0 &  1 & -1 &  1 & -1\\		
		0 &  0 &  1 &  0 &  1 &  1 &  1 &  1 &  1\\		
		0 &  0 &  0 &  0 &  0 &  1 &  1 & -1 & -1\\
		0 &  0 &  0 &  0 &  0 &  1 & -1 & -1 &  1\\
		0 &  0 &  0 &  0 &  0 &  1 &  1 &  1 &  1\\
	\end{pmatrix},
\end{equation}
we can obtain $s_g^1=s_g^2=s_1$, $s_g^3=s_g^4=s_g^5=s_2$, $H=\frac{s_g^0s_g^3-4s_g^0s_g^8+3s_g^3s_g^8}{3s_g^0s_g^3s_g^8}$, $J=\frac{s_g^3+3}{3c^2s_g^3}$, $K=\frac{1}{3c^2}$, and the following moments can be obtained, 
\begin{equation}\label{MgMF2}
	\mathbf{m}_g=\begin{pmatrix}
		\rho_0\\
		\frac{\rho u}{c}\\
		\frac{\rho v}{c}\\
		\frac{p+um_x^{\phi C}+\rho uu}{c^2}\\
		\frac{um_y^{\phi C}+vm_x^{\phi C}+2\rho uv}{2c^2}\\
		\frac{p+vm_y^{\phi C}+\rho vv}{c^2}\\
		\frac{\rho v}{3c}\\
		\frac{\rho u}{3c}\\
		\frac{p+\mathbf{u}\cdot\mathbf{m}^{\phi C}+\rho\mathbf{u}\cdot\mathbf{u}}{3c^2}\\
	\end{pmatrix},\quad
	\mathbf{m}_F=\begin{pmatrix}
		\mathbf{u}\cdot\nabla\rho\\
		\frac{F_x}{c}\\
		\frac{F_y}{c}\\
		\frac{2F_xu+\partial_t\left(um_x^{\phi C}\right)}{c^2}+\frac{3u\partial_x\rho+v\partial_y\rho}{3}\\
		\frac{u\partial_y\rho+v\partial_x\rho}{3}+\frac{\partial_t\left(um_y^{\phi C}+vm_x^{\phi C}\right)+2\left(uF_y+vF_x\right)}{3c^2}\\
		\frac{2F_y+v\partial_t\left(vm_y^{\phi C}\right)}{c^2}+\frac{u\partial_x\rho+3v\partial_y\rho}{3}\\
		\frac{F_y}{3c}\\
		\frac{F_x}{3c}\\
		\frac{\partial_t\left(\mathbf{u}\cdot\mathbf{m}^{\phi C}\right)+2\mathbf{u}\cdot\mathbf{F}}{3c^2}+\frac{\mathbf{u}\cdot\nabla\rho}{3}\\
	\end{pmatrix}.
\end{equation}

Similarly, we can also calculate the moments in phase field,

\noindent CH equation:
\begin{equation}\label{MfMRCH2}
	\begin{aligned}
		&\mathbf{m}_f=\left(\phi,\frac{\phi u}{c},\frac{\phi v}{c},\frac{\eta\mu_{\phi}}{3},0,\frac{\eta\mu_{\phi}}{3},\frac{\phi v}{3c},\frac{\phi u}{3c},\frac{\eta\mu_{\phi}}{9}\right)^T,\\
		&\mathbf{m}_R=\left(0,\frac{\partial_t\left(\phi u\right)}{c},\frac{\partial_t\left(\phi v\right)}{c},0,0,0,\frac{\partial_t\left(\phi v\right)}{3c},\frac{\partial_t\left(\phi u\right)}{3c},0\right)^T,
	\end{aligned}	
\end{equation}

\noindent AC equation:
\begin{equation}\label{MfMRAC2}
	\begin{aligned}
		&\mathbf{m}_f=\left(\phi,\frac{\phi u}{c},\frac{\phi v}{c},\frac{\eta\phi}{3},0,\frac{\eta\phi}{3},\frac{\phi v}{3c},\frac{\phi u}{3c},\frac{\eta\phi}{9}\right)^T,\\
		&\mathbf{m}_R=\left(0,\frac{c\lambda n_x}{3}+\frac{\partial_t\left(\phi u\right)}{c},\frac{c\lambda n_y}{3}+\frac{\partial_t\left(\phi v\right)}{c},0,0,0,\frac{c\lambda n_y}{9}+\frac{\partial_t\left(\phi v\right)}{3c},\frac{c\lambda n_x}{9}+\frac{\partial_t\left(\phi u\right)}{3c},0\right)^T.
	\end{aligned}
\end{equation}

We will use the first orthogonal transformation matrix $\mathbf{M}$ [Eq.\,(\ref{M1})] and the corresponding moments in our numerical simulations. 
In addition, it should be noted that the derivative terms in the present LB method should be discretized with suitable difference schemes. For simplicity, the explicit Euler scheme $\partial_{t}\chi\left(t\right)=\left[\chi\left(t\right)-\chi\left(t-\Delta t\right)\right]/\Delta t$ is adopted for the temporal derivatives in $\mathbf{m}_F$ and $\mathbf{m}_R$, and the second-order isotropic central schemes are applied for the gradient and Laplacian operators \cite{Guo2011PRE,Lou2012EPL},
\begin{equation}\label{central1}
	\nabla\chi\left(\mathbf{x}\right)=\sum_{i\neq0}\frac{\omega_i\mathbf{c}_i\chi\left(\mathbf{x}+\mathbf{c}_i\Delta t\right)}{c_s^2\Delta t},
\end{equation}
\begin{equation}\label{central2}
	\nabla^2\chi\left(\mathbf{x}\right)=\sum_{i\neq0}\frac{2\omega_i\left[\chi\left(\mathbf{x}+\mathbf{c}_i\Delta t\right)-\chi\left(\mathbf{x}\right)\right]}{c_s^2\Delta t^2}.
\end{equation}

We would also like to point out that in the framework of LB method for phase-field AC equation, one can also obtain the local scheme for the gradient of the order parameter and its gradient norm \cite{Wang2016PRE},
\begin{subequations}\label{local}
	\begin{equation}
		|\nabla\phi|=\frac{-|C|-B}{A},
	\end{equation} 
	\begin{equation}
		\nabla\phi=\frac{C}{A+B/|\nabla\phi|},
	\end{equation}
\end{subequations}
where $A=-M_{\phi}-0.5c_s^2\Delta t$, $B=M_{\phi}\lambda$, and $C=\sum_{i}\mathbf{c}_if_i-\phi\mathbf{u}+0.5\Delta t\partial_{t}\left(\phi\mathbf{u}\right)$.

%%%% Bibliography  %%%%%%%%%%
\bibliographystyle{elsarticle-num} 
\bibliography{reference}
\end{document}